\documentclass[12pt]{article}

\usepackage{graphicx}
\usepackage{amsmath,amssymb}
\usepackage{bm}
\allowdisplaybreaks
\begin{document}
\baselineskip=0.7cm
\renewcommand{\figurename}{Fig.\@}
\renewcommand{\thesection}{\arabic{section}.}
\renewcommand{\theequation}{\arabic{section}.\arabic{equation}}
\renewcommand{\thesubsection}{\arabic{section}.\arabic{subsection}}
\makeatletter
\def\section{\@startsection{section}{1}{\z@}{-3.5ex plus -1ex minus 
 -.2ex}{2.3ex plus .2ex}{\large}} 
\def\subsection{\@startsection{subsection}{2}{\z@}{-3.25ex plus -1ex minus 
 -.2ex}{1.5ex plus .2ex}{\normalsize\it}}
\makeatother
\makeatletter
\def\lesim{\mathrel{\mathpalette\gl@align<}}
\def\gtsim{\mathrel{\mathpalette\gl@align>}}
\def\gl@align#1#2{\lower.7ex\vbox{\baselineskip\z@skip\lineskip.2ex%
  \ialign{$\m@th#1\hfil##\hfil$\crcr#2\crcr\sim\crcr}}}
\makeatother

\makeatletter
\def\@cite#1#2{$^{\hbox{\scriptsize{#1\if@tempswa , #2\fi})}}$}
\makeatletter

\def\thefootnote{\alph{footnote}}


\newcommand{\sn}{{\rm\,sn\,}}
\newcommand{\cn}{{\rm\,cn\,}}
\newcommand{\dn}{{\rm\,dn\,}}

\vspace*{3cm}
\begin{center}
\Large 
Generalized Hamilton-Jacobi theory of Nambu Mechanics
\vspace{0.7cm}

\normalsize
 \vspace{0.4cm}
Tamiaki {\sc  Yoneya} \footnote{
Emeritus Professor}

\vspace{0.3cm}

{\it Institute of Physics, University of Tokyo\\
Komaba, Meguro-ku, Tokyo 153-8902}

\vspace{1cm}
Abstract
\end{center}
\vspace{0.4cm}
We develop a Hamilton-Jacobi-like formulation 
of Nambu mechanics. 
The Nambu mechanics, originally proposed by Nambu more than four decades ago, provides a remarkable extension of the standard Hamilton equations of motion 
in even-dimensional phase space with a single Hamiltonian to 
a phase space of three
 (and more generally,  arbitrary)  
dimensions with two Hamiltonians ($n$ Hamiltonians 
in the case of $(n+1)$-dimensional phase space) from the viewpoint of the Liouville theorem. However,  
it has not been formulated seriously in the spirit of Hamilton-Jacobi theory. 
The present study is motivated to suggest 
a possible direction towards quantization from a new perspective. 

\newpage
\section{Introduction}
In 1973, Nambu\cite{nambu1} proposed the following system of equations of motion for the flows of 
a point $(\xi^1,\xi^2,\xi^3)$  
in a three-dimensional phase space $\mathbb{R}^3$:
\begin{align}
\frac{d\xi^i}{dt}=\{H, G, \xi^i\}\equiv X^i, 
\label{nambueq}
\end{align}
where the bracket notation (the Nambu bracket) 
on the r.h.side is defined for an 
arbitrary triplet of three functions $(K,L,M)$ on the phase space in terms of 
three-dimensional Jacobian,\footnote{We assume the usual summation convention 
for coordinate indices in phase space.}
\begin{align}
\{K, L, M\}=\frac{\partial(K,L,M)}{\partial(\xi^1,\xi^2,\xi^3)}
=\epsilon^{ijk}\partial_iK\partial_jL\partial_kM.
\end{align}
Therefore the vector field $X^i$ defined in \eqref{nambueq}, generating the lines of flows in the phase space, is equal to
\begin{align}
X^i=\epsilon^{ijk}\partial_jH\partial_kG, 
\end{align}
in which two conserved functions $H$ and $G$, or ``Hamiltonians", 
govern 
the time evolution on an equal footing. 
The phase-space coordinates $\xi^i$ satisfy a 
``canonical" Nambu bracket relation,
\begin{align}
\{\xi^i,\xi^j, \xi^k\}=\epsilon^{ijk}.
\label{canob}
\end{align}
Thus we have a natural extension of the Hamilton 
equations ($i=1,\ldots, n$) of motion,  
\begin{align}
&\frac{dq^i}{dt}=\frac{\partial H}{\partial p_i}=\{H, q^i\}, 
\label{Heq1}\\
&\frac{dp_i}{dt}=-\frac{\partial H}{\partial q^i}=\{H, p_i\}, 
\label{Heq2}
\end{align}
 in ordinary phase spaces, which are 
intrinsically of even ($2n$) dimensions; we are liberated 
from the restriction of paired sets $(p_i, q^i)$ of independent canonical variables, satisfying the canonical Poisson bracket relations 
$\{p_i, q^j\}=\delta_i^j$. 

Once this generalization is given, it is obvious that 
similar systems in the phase spaces of arbitrary dimensions 
$n+1$,  
irrespectively of odd or even dimensions, can be constructed by 
replacing 3-dimensional Jacobian in \eqref{nambueq} by 
a general $(n+1)$-dimensional Jacobian with $n$ Hamiltonians. 
Let us call this general case the Nambu mechanics of order $n$. 

The proposal would be quite suggestive from the standpoint of exploring 
new methods to express fundamental 
dynamical laws. 
Nambu himself was motivated by the Liouville theorem 
of ordinary Hamilton mechanics ($
\partial_iX^i=0$), aiming at an 
extension of statistical mechanics. 
He stressed in particular that the Euler equations of motion 
for a free rigid rotator can be cast in the form \eqref{nambueq}
by identifying the component $\ell_i$ of angular momentum
 in the 
body-fixed frame to be the canonical 
coordinates $\xi^i$, with two Hamiltonians, $G=\frac{1}{2}\bigl(
(\xi^1)^2/I_1+(\xi^2)^2/I_2+(\xi^3)^2/I_3\bigr)$ and $
H=\frac{1}{2}\bigl((\xi^1)^2+(\xi^2)^2+(\xi^3)^2\bigr).$

During the first two decades after its proposal, this remarkable 
concept had been a matter of 
interest only for a relatively small circle of mathematical physicists. 
An important basis for further developments 
was laid, among others, by Takhtajan\cite{takh}, who found a crucial identity clarifying the canonical structure of Nambu mechanics, 
now called the fundamental identity (FI) for the 
Nambu bracket, and 
also formulated an action principle. 
In the sequel to this long initial period, we gradually came to recognize the relevance of these ideas to physics, especially, to string/membrane theory (or M-theory),\footnote{For these developments, interested readers can refer to Refs. {\it e.g.} \cite{yone} and 
\cite{homatsu}, and references therein. } and the interests in Nambu mechanics and Nambu bracket have been renewed and broadened 
fruitfully in the past two decades. 
The purpose of the present paper, however, is not to pursue further 
such new directions for applications, but rather to fill a 
missing aspect of classical Nambu mechanics by going back to the 
spirit of the original proposal. 

What we 
address is whether and how some sort of Hamilton-Jacobi (HJ)-like formalism is possible in this system. To the author's knowledge, 
this question has never been pursued in appropriate depth in the past literature.\footnote{The only work of which the present 
author is aware in connection with this is Ref. \cite{esta} where 
the problem of HJ theory in Nambu mechanics is briefly mentioned without 
any concrete formulation. } 
From a purely technical point of view, the system of the 
equations \eqref{nambueq} of order $n$ may 
be regarded as a special new class of integrable systems in which 
we have $n$ independent conserved quantities for 
$n+1$ variables obeying ordinary 
differential equations of first order with 
respect to time. The latter property ensures 
that we can solve the system, in principle, by reducing it to a single 
quadrature directly at the level of the equations of motion, without higher 
apparatus such as HJ-like formalism. Conceptually, however, 
we can take a different attitude. The whole endeavors of physicists for 
developing Nambu mechanics have been devoted to 
uncovering possible deeper structures behind its surface, as 
suggested, e. g., from its stringent and higher 
symmetry properties, 
in hopes of utilizing them for fundamental physics. 
From this viewpoint, it must be worthwhile 
exploring the possibility of HJ-like formalism by focusing 
its implications towards quantization of Nambu mechanics, 
a subject that has been quite elusive even to this day. 
We 
should recall here the well-known significance that classical HJ theory had given 
to the creation of quantum mechanics. This is not a mere 
accident: it reflects the important fact that 
the idea of the HJ formalism already contains some 
essential elements of quantum mechanics. 

In fact, it turns out that 
the procedure towards an HJ-like formulation of Nambu mechanics 
is not straightforward. 
To achieve our goal, it will be 
useful to reformulate the basic ideas of the HJ formalism in 
ordinary Hamilton mechanics in a manner 
that does not rely too much on 
the standard textbook formulation. 
For this reason and also for the purpose of making 
the present paper accessible as widely as possible to readers of various backgrounds  
in a reasonably self-contained manner, 
we start in the next preliminary section by recounting 
ordinary HJ theory from a slightly nonstandard but physical viewpoint. 
That will guide us in constructing HJ-like formulations 
of Nambu mechanics in Sects. 3 and 
4, where we 
restrict ourselves to the simplest case of 
three-dimensional ($n=2$) phase space. 
In Sect. 5, 
as an application of our formalism, 
we demonstrate some 
concrete computations to solve our generalized 
HJ equations and to rederive the Nambu equations of motion from them, taking the example of 
the (free) Euler top. In Sect. 6,  
we propose a possible new standpoint towards quantization, 
on the basis of our main results.  
Appendix A 
is devoted to a discussion on the extension 
of the present formalism to general Nambu mechanics of 
higher orders.

\section{Some preliminaries}
\setcounter{equation}{0}

Let us first recall briefly what the standard HJ theory is. 
The usual textbook accounts presuppose the 
existence of an action functional and also the general formula of finite canonical 
transformations in terms of the action function $S(q,t;Q)$ as a 
generating function defined 
on the configuration space of generalized coordinates $q^i,\, (i=1, \ldots, n)$ adjoined with 
time $t$. 
The ``action function" means that the action integral 
as a functional of arbitrary trajectories is now 
evaluated for particular trajectories that 
solve the equations of motion, by specifying an arbitrary point $q^i$ at time $t$ as the endpoint condition, together 
with the initial conditions, $q^i=Q^i$, at $t=0$. If we fix the 
$Q^i$ as constants, the solutions of the equations of motion, 
Eqs. \eqref{Heq1} and 
\eqref{Heq2}, are uniquely fixed 
since we now have $2n$ conditions. These conditions in turn 
determine, at least locally, the action as a function (or a field), $S(q,t; Q)$, of $n$ independent 
variables $q^i$ with fixed parameters $Q^i$'s. 
Then, the momenta as the canonical conjugates of the 
$q^i$ and hence the 
Hamiltonian $H=H(p,q)$ also become 
fields, and are expressed in terms of the 
action function as 
\begin{align}
&p_i(q,t)=\frac{\partial S}{\partial q^i}, \label{pfield}\\
&H(q, p(q,t))+\frac{\partial S}{\partial t}=0, \label{oHJ}
\end{align}
which constitute the HJ equation as a partial 
differential equation of first order for the action function or field $S$. 
These equations are usually derived on the basis of familiar 
variational principles associated with the action integral. 
The parameters $Q^i$ emerge as constants of integration 
for the HJ equation; such a solution with $n$ independent 
integration constants is called a ``complete solution". 
Given a complete solution, 
we obtain general 
solutions, now with $2n$ integration constants, of the Hamilton equations of motion 
by simple quadratures, {\it after} imposing on it the 
conditions (called ``Jacobi conditions" for convenience in the present paper), 
\begin{align}
\frac{\partial S}{\partial Q^i}=-P_i, 
\label{jacobi}
\end{align}
by introducing the $P_i$ as new and additional constant parameters, and 
solving them for $q^i=q^i(t; P, Q)$ and $p_i=p_i(t;P, Q)$. 
The solvability is guaranteed by requiring
\begin{align}
{\rm det}\Bigl(\frac{\partial^2S}{\partial Q^i\partial q^j}\Bigr)\ne 0. 
\label{solv}
\end{align}
Equation \eqref{pfield} together with their counterpart equations \eqref{jacobi} at $t=0$ 
are interpreted as defining relations for a 
canonical transformation from $(p, q;t)$ to $(P,Q;0)$ which 
unfolds the time development by sending 
$H$ to zero as signified by \eqref{oHJ}. 
Finally, condition \eqref{jacobi} 
can be rephrased in the following form. 
Since the HJ equation involves the action field through its 
derivatives, we can always shift it by adding a constant. 
In particular, by choosing a shift in the form 
$S\rightarrow \tilde{S}=S+P_iQ^i$, condition \eqref{jacobi} 
is $\partial \tilde{S}/\partial Q_i=0$. This interpretation 
is appropriate to Schr\"{o}dinger's wave mechanical  quantization: the condition $\partial \tilde{S}/\partial Q_i=0$ naturally arises in deriving classical trajectories in the limit $\hbar\rightarrow 0$ from a 
wave function whose phase is  $\tilde{S}/\hbar$.

Now, in the case of Nambu mechanics, we have an action functional 
proposed in Ref.\cite{takh}. However, it is a functional {\it not} 
of one-dimensional trajectories, but of $n$-dimensional 
(hyper) surfaces, represented say by $\xi^i(t, s_1,\ldots, s_{n-1})$, which consist of continuous families of one-dimensional trajectories, 
parametrized by spatial world-hypersurface coordinates $(s_1,\ldots, s_{n-1})$, as if we were treating objects extending in $(n-1)$ dimensions. 
This is not convenient for our purpose. 
Although nothing prevents us from treating arbitrary 
families of trajectories, such an approach forces us 
to introduce too many inessential and unphysical 
degrees of freedom, caused by the presence of 
the additional parameters $s_i$ that are {\it not} necessary for 
describing the true dynamical 
degrees of freedom: $s_i$ are simply redundant, at least for the present 
purpose, since 
they do not correspond to any ``energetic" couplings among trajectories, 
a property that is related to the existence of 
$n$ independent ``Hamiltonians", in spite of the fact that 
we have only a single time for the dynamical evolution. 

To make things worse, 
we do not know an appropriate and useful
characterization for {\it finite} canonical transformations,  
 to a similar extent that we are familiar with in 
ordinary Hamilton mechanics. 
The origin of this difficulty will be discussed in section 3. 
Thus it is not at all straightforward to 
proceed if we try to 
mimic the above procedure. We therefore start with 
a different root, due originally to Einstein (Ref. \cite{einstein}), 
 that does not presuppose 
any knowledge of an action functional, nor of canonical transformations.\footnote{{\it Historical remark}: Einstein's original intention was 
to extend the Sommerfeld quantum condition to 
non-separable cases, giving a coordinate-independent formulation of the semiclassical 
quantum condition, the significance of which is now well known in connection 
with the theory of quantum chaos. 
This work (the second of Ref.\cite{einstein}) played an influential role 
in a forming period of quantum mechanics and was 
cited by Schr\"{o}dinger and also by 
de Broglie in their monumental works.  In this attempt, Einstein gave a simple descriptive formulation of the HJ formalism, which is, 
according to him, ``free of surprising tricks of the trade".\cite{einstein}    
Unfortunately, this small but useful observation is almost 
forgotten now. The present author could not find any appropriate 
reference that explicitly mentioned his observation.  }


Since, under the above conditions on the initial points and endpoints,  
the trajectories are uniquely determined, we can follow the time 
development of the $p_i$ as functions the $q^i$ on the configuration space, $p_i=p_i(q,t)$. 
Then we rewrite the l.h.side of \eqref{Heq2} as
\[
\frac{\partial p_i}{\partial t}+\frac{\partial p_i}{\partial q^j}
\frac{dq^j}{dt}=\frac{\partial p_i}{\partial t}
+\frac{\partial H}{\partial p_j}\frac{\partial p_i}{\partial q^j},
\]
by using \eqref{Heq1}. We then obtain 
\begin{align}
\frac{\partial p_i}{\partial t}+\frac{\partial H}{\partial q^i}
+\frac{\partial H}{\partial p_j}\frac{\partial p_i}{\partial q^j}=0. 
\label{EE}
\end{align}
These are partial differential equations of first order 
for the vector field 
$p_i(q,t)$ on the configuration space of independent variables $(q,t)$, 
since $\partial H/\partial p_i$ and $\partial H/\partial q^i$ are 
known algebraic functions of the canonical variables. We call this type of partial differential equation an 
``Euler-Einstein (EE) equation" for later convenience, since they are 
analogous to the standard Euler equations in fluid mechanics 
where the role of the vector field $p_i$ is played by the 
velocity field of the fluid. 

Here, following Einstein, it is convenient to introduce the 
notation
\begin{align}
\bar{H}=\bar{H}(q,t)=H(p(q,t),q), \nonumber 
\end{align}
which helps us to make clear the difference in independent 
variables between $H$ and $\bar{H}$, before performing 
partial differentiation. Now let us further require that the motion 
of fluid has no vorticity:
\begin{align}
\frac{\partial p_j}{\partial q^i}-\frac{\partial p_i}{\partial q^j}=0, 
\nonumber 
\end{align}
which guarantees the existence of the ``velocity" potential $J$ 
such that
\begin{align}
p_i=\frac{\partial J}{\partial q^i}. \nonumber 
\end{align}
Then the sum of the
second and third terms in the l.h.side of \eqref{EE} 
simply 
takes the form 
$\partial{\bar{H}}/\partial q_i$. 

The EE equations now take the form, 
\[
\frac{\partial}{\partial q^i}\Bigl(
\frac{\partial J}{\partial t}+\bar{H}\Bigr)=0.
\nonumber 
\]
Thus we arrived at a single equation
\[
\frac{\partial J}{\partial t}+\bar{H}=f(t),
\]
where $f$ is an arbitrary function of time only. 
Obviously, the arbitrariness of $f$ does not affect the dynamics, since 
we can always redefine a new potential function $S$ such that 
$\partial J/\partial t-f=\partial S/\partial t$ and 
$p_i=\partial S/\partial q_i$. Hence we obtain the 
HJ equation
\begin{align}
\frac{\partial S}{\partial t}+\bar{H}=0. 
\label{oHJ2}
\end{align}

Conversely, we can also reproduce the original Hamilton 
equations of motion, in a manner analogous to the 
way we make transitions 
from the Euler picture to the Lagrange picture in fluid mechanics. 
Suppose we know the trajectories in configuration space as functions, $q^i=q^i(t)$, satisfying 
\eqref{Heq1}. 
Then the motions of the momenta as functions of time automatically  satisfy 
\eqref{Heq2}  as a consequence of the HJ equation: 
\[
\frac{dp_i}{dt}=\frac{\partial p_i}{\partial t}+
\frac{\partial p_i}{\partial q^j}\frac{dq^j}{dt}=\frac{\partial^2S}{\partial q^i\partial t}+\frac{\partial^2S}{\partial q^j\partial q^i}
\frac{dq^i}{dt}=-\frac{\partial H}{\partial q^i}
-\frac{\partial H}{\partial p_j}\frac{\partial^2S}{\partial q^i\partial q^j}
+\frac{\partial^2S}{\partial q^j\partial q^i}
\frac{dq^i}{dt}=-\frac{\partial H}{\partial q^i}.
\] 
This can be regarded as a version of the integrability condition 
for the Hamilton-Jacobi equation if $p_i=\partial S/\partial q_i$ is 
separately treated as differential equations. Its validity is actually guaranteed by our derivation, since 
this calculation merely inverts the process from Eq. \eqref{Heq2} to Eq.
\eqref{oHJ2}. This is one of the merits of the Einstein 
approach.

On the other hand, Eq. \eqref{Heq1} itself is obtained 
by the Jacobi condition \eqref{jacobi}, given a complete solution to 
\eqref{oHJ2}: by a total differentiation of \eqref{jacobi} by $t$, 
we obtain
\begin{align}
0=\frac{\partial^2S}{\partial Q^i\partial t}+\frac{\partial^2S}{\partial Q^i\partial q^j}\frac{dq^j}{dt}=-\frac{\partial \bar{H}}{\partial Q^i}
+\frac{\partial^2S}{\partial Q^i\partial q^j}\frac{dq^j}{dt}
=-\frac{\partial H}{\partial p_j}\frac{\partial^2S}{\partial q^j\partial Q_i}
+\frac{\partial^2S}{\partial Q^i\partial q^j}\frac{dq^j}{dt}, 
\nonumber 
\end{align}
which gives Eq. \eqref{Heq1} under condition \eqref{solv}.  

The above procedures actually fit well into an abstract but modern language of differential forms. 
First define a closed (and exact) 2-form in the phase space $(p, q, t)$ adjoined by a time variable, 
\begin{align}
\omega^{(2)}=dp_i\wedge dq^i-dH\wedge dt =
d\omega^{(1)}
\end{align}
where
\begin{align}
\omega^{(1)}=p_idq^i-Hdt. 
\end{align}
The EE equations together with the 
vortex-free condition are equivalent to a demand that the 
2-form $\omega^{(2)}$ vanishes when it is evaluated 
under the projection to the configuration space $(q^i,t)$, by assuming $p_i=p_i(q,t)$ and hence $dp_i=\frac{\partial p_i}{\partial q^j}dq^j
+\frac{\partial p_i}{\partial t}dt$:
\begin{align}
\bar{\omega}^{(2)}\equiv \omega^{(2)}|_{(q,t)}=\frac{1}{2}\Bigl(
\frac{\partial p_j}{\partial q^i}
-\frac{\partial p_i}{\partial q^j}\Bigr)dq^i\wedge dq^j
-\Bigl(\frac{\partial p_i}{\partial t}+\frac{\partial \bar{H}}{\partial q^i}\Bigr) dq^i\wedge dt=0. 
\end{align}
Obviously, the vortex-free condition is nontrivial only 
for $n\ge 2$. 
The requirements of vortex-free flow and consequently 
of the HJ equation for potential function $S$ as the 
vanishing condition for $\bar{\omega}^{(2)}$ are formulated 
equivalently to the condition that the 1-form $\omega^{(1)}$ in the phase space becomes exact {\it after} the projection to the configuration space: namely, the equality 
\begin{align}
\bar{\omega}^{(1)}\equiv \omega^{(1)}|_{(q,t)}=dS=\frac{\partial S}{\partial q^i}dq^i+\frac{\partial S}{\partial t}dt
\end{align}
 is nothing but the HJ equation.

It is also useful, though not 
essential to our development, to note the following in understanding the connection 
of $\omega^{(2)}$ to the action principle. 
If we do not make the projection by treating 
$p_i$ and $q^j$ as independent variables, we can characterize it by (Ref. \cite{arnold})
\begin{align}
i_{\tilde{V}}(\omega^{(2)})
=0, 
\label{null}
\end{align}
where the symbol $i_{L}({\cdot})$ denotes in general the operation of internal multiplication of a vector differential operator $L$ on differential forms abbreviated as ``$\, \cdot"\,$. 
In the present case,  
\begin{align}
L=\tilde{V}=V+\frac{\partial}{\partial t}, \quad V\equiv \frac{\partial H}{\partial p_i}\frac{\partial}{\partial q^i}
-\frac{\partial H}{\partial q^i}\frac{\partial}{\partial p_i}, 
\end{align}
corresponding to the Hamilton equations of motion. In this sense, 
$\tilde{V}$, whose component form is 
$(-\partial H/\partial q^i, \partial H/\partial p_i, 1)$, is called a ``null vector" for the form $\omega^{(2)}$. 
This property is akin to the above formulation with projection. 
The null condition \eqref{null}, which 
is an analogue of our vanishing condition, is essentially 
equivalent, due to the Stokes theorem,  to
 the usual variational principle for 
the action integral $S[p,q;t]=\int \omega^{(1)}$ in the phase space.  
A null vector in this sense is often called a ``line field", alternatively, 
reflecting the fact that its properties are similar to those 
associated with the 
Faraday magnetic lines of forces in electromagnetism: the null 
condition is analogous to an obvious property that 
the circulation or rotation (corresponding to 
$\omega^{(2)}$) of vector potential (corresponding to 
$\omega^{(1)}$)
along the boundary of an infinitesimal square containing magnetic lines of forces is always zero. For more details about this, see Ref. \cite{arnold}. A moral here is that the HJ theory 
can also be interpreted as a counterpart of the null 
condition \eqref{null} under the Einstein projection from 
phase space to the base configuration space. 
Since the null condition in phase space includes the 
action principle as one of its consequences, the vanishing condition can 
be regarded as 
amounting to replacing the action principle without the  action integral explicitly.

For guiding our later development, it is convenient to regard the 
HJ theory 
as consisting of three steps. We call the initial process to obtain 
the EE equations from the equations of motion ``step I". 
Here, in terms of the 
language of fiber bundles, we first decompose phase space into 
the configuration space as the base space and consider the space parametrized 
by the momentum vector $p_i$ (cotangent 
vector) as the fiber lying on each point of the base space. 
In this sense, the ordinary phase space is usually called the ``cotangent bundle". 
In step I, we describe the motions in phase space by 
regarding the section specified by functions $p_i=p_i(q,t)$ as a 
 dynamical (covariant) vector field on the base space. 
The next step from the EE equations to the HJ equation by 
demanding the vanishing of the 2-form under the projection is called ``step II". 
The remaining and final step in which we re-derive the equations 
of motion from the HJ equations is called ``step III".  
Here we constrain complete solutions for the HJ 
equations by designing a device with a particular (Jacobi) prescription 
through which we can determine the trajectories with respect to the base space coordinates directly 
as functions of time. 
It will turn out that this step, being actually a 
decisive part of the HJ formalism as 
a preliminary to quantization, is a hurdle 
 in establishing 
the generalized HJ formalism for Nambu mechanics. 

With this basic understanding of the nature of HJ formalism, we can now proceed to our main objective, the construction 
of an HJ-like formalism for Nambu mechanics.  
First we have to determine the 
decomposition of the phase space $(\xi^1,\xi^2, \xi^3)$ 
into a base space and fibers lying on it. Initially, there are two 
possibilities: designating the dimensions 
of fibers and the base spaces by a symbol (fiber/base), 
\begin{align}
&(1/2)\mbox{ decomposition}: (\xi^1, \xi^2, \xi^3) 
\rightarrow \bigl(\xi^1, \xi^2, \xi^3(\xi_1,\xi_2, t)\bigr),\nonumber \\
&(2/1) \mbox{ decomposition}: (\xi^1, \xi^2, \xi^3) 
\rightarrow \bigl(\xi^1,\xi^2(\xi^1,t), \xi^3(\xi_1, t)\bigr), 
\nonumber  
\end{align}
where, in the first case, the base space has two dimensions 
with coordinates $(\xi^1,\xi^2)$ and the fibers 
are one-dimensional spaces of $\xi^3$, whose sections 
are described by one component field $\xi^3(\xi^1,\xi^2,t)$. 
In the second case, the base space is just a line 
whose coordinate is $\xi^1$, and the fibers have two dimensions whose 
sections are described by two-component 
fields $\bigl(\xi^2(\xi^1,t), \xi^3(\xi^1,t)\bigr)$. 
From the viewpoint of the equations of motions whose solutions 
can be specified by three independent parameters, 
the first case corresponds to specifying the endpoint (at the time $t$) of 
the trajectories in the base-configuration space at 
$(\xi^1,\xi^2)$ and the initial condition is 
assigned a single implicit parameter, say $Q_1$, such that the trajectories 
are uniquely determined, in principle, and the sections of the one-dimensional fibers are described by 
$\xi^3$ as a field $\xi^3(\xi^1,\xi^2,t; Q_1)$. 
In the second case, we specify the endpoint at 
$\xi^1$ on the one-dimensional base space, and 
the initial condition by two implicit constant parameters, say $(Q_1, Q_2)$.\footnote{For the integration constants of Nambu mechanics, no discrimination is assigned regarding the upper (contravariant) or lower (covariant) positions 
of their indices.}
 Then the two-component field $\bigl(\xi^2(\xi^1,t;Q_1,Q_2), 
\xi^3(\xi^1,t;Q_1,Q_2)\bigr)$ gives a section of the 
two-dimensional fibers. We stress that, apart from questions on the 
global existence of these functions, the odd dimensionality 
of phase space 
does not obstruct us at all: in the present approach, the specification of the conditions for 
determining trajectories uniquely is most essential and sufficient 
for our purpose, irrespectively 
of the dimensions of phase space. 

\section{(1/2)-formalism}
\setcounter{equation}{0}
\subsection{Steps I and II}
Let us start from the case of (1/2) decomposition. 
We define
\begin{align}
&\bar{H}=\bar{H}(\xi^1,\xi^2, t)=H(\xi^1,\xi^2, \xi^3(\xi^1,\xi^2,t)),\\
&\bar{G}=\bar{G}(\xi^1,\xi^2, t)=G(\xi^1,\xi^2, \xi^3(\xi^1,\xi^2,t)).
\end{align}
The EE equation for the field $\xi^3=\xi^3(\xi^1,\xi^2,t)$ is derived by following the procedures  
explained in the previous section. The partial 
differentiations will be abbreviated as 
$\frac{\partial}{\partial \xi^i}=\partial_i,  \frac{\partial}{\partial t}=\partial_t=\partial_0$. We first have
\begin{align}
\frac{\partial(H,G)}{\partial(\xi^1,\xi^2)}=\frac{d\xi^3}{dt}=\partial_t \xi^3+\partial_i\xi^3
\frac{d\xi^i}{dt}=\partial_t\xi^3+\partial_1\xi^3
\frac{\partial(H,G)}{\partial(\xi^2,\xi^3)}+
\partial_2\xi^3\frac{\partial (H,G)}{\partial(\xi^3,\xi^1)}.
\label{xi3eq}
\end{align}
Here and throughout this section 
it should be understood, unless stated otherwise,  
that, for $(H,G)$ {\it without} bars,  
$\xi^3=\xi^3(\xi^1,\xi^2,t)$ is substituted after operating 
partial differentiations by treating all three 
coordinates of the phase space independently: 
e. g., 
$\partial_i\bar{H}=\partial_iH+\partial_3H
\partial_i\xi^3$. Thus we have
\begin{align}
\frac{\partial(H,G)}{\partial(\xi^2,\xi^3)}&
=(\partial_2\bar{H}-\partial_3H\partial_2\xi^3)\partial_3G
-(\partial_2\bar{G}-\partial_3G\partial_2\xi^3)\partial_3H=\partial_2\bar{H}\partial_3G-\partial_2\bar{G}\partial_3H,\nonumber \\
\frac{\partial(H,G)}{\partial(\xi^3,\xi^1)}&=
\partial_3H(\partial_1\bar{G}-\partial_3G\partial_1\xi^3)-
\partial_3G(\partial_1\bar{H}-\partial_3H\partial_1\xi^3)=\partial_3H\partial_1\bar{G}
-\partial_3G\partial_1\bar{H},\nonumber \\
\frac{\partial(H,G)}{\partial(\xi^1,\xi^2)}
&=\partial_1\bar{H}\partial_2\bar{G}-\partial_1\bar{G}\partial_2
\bar{H}-\partial_1\xi^3(\partial_3H\partial_2\bar{G}
-\partial_3G\partial_2\bar{H})-\partial_2\xi^3
(\partial_1\bar{H}\partial_3G
-\partial_1\bar{G}\partial_3H).\nonumber 
\end{align}
When these results are put together, 
we find that the terms proportional to $\partial_i\xi^3 \, (i=1,2)$ 
cancel between r.h. and l.h. sides of \eqref{xi3eq}, 
and we obtain
\begin{align}
\partial_t\xi^3=\partial_1\bar{H}\partial_2\bar{G}-\partial_1\bar{G}
\partial_2\bar{H}=\partial_1(\bar{H}\partial_2\bar{G})-
\partial_2(\bar{H}\partial_1\bar{G}).
\label{1/2-EE}
\end{align}
Since the r.h.side is 
a known (algebraic) function of $\xi^1, \xi^2, \xi^3(\xi^1,\xi^2,t)$ and $ \partial_i\xi^3(\xi^1,\xi^2,t)$, 
this is the desired EE equation. We have completed step I.

There is now a natural way to step II. 
The form \eqref{1/2-EE} exhibited in the last 
equality suggests itself to represent, without losing generality, 
the field $\xi^3$ as the vorticity of a 
two-component vector field $(S_1,S_2)$ as
\begin{align}
\xi^3\equiv \epsilon^{3ij}\partial_iS_j=\partial_1S_2-\partial_2S_1. 
\label{1/2-HJ0}
\end{align}
Then, Eq. \eqref{1/2-EE} can be expressed as (paired) partial differential 
equations for $S_i$, being supplemented with an arbitrary 
scalar field $S_0$ that does not contribute to the 
vorticity of $\partial_tS_i$ on the l.h. side of Eq. \eqref{1/2-EE}:
\begin{align}
&\partial_tS_i
=\bar{H}\partial_i\bar{G}+\partial_iS_0 \label{1/2-HJ}\\
\bar{H}=H(\xi^1,\xi^2, &\partial_1S_2-\partial_1S_2), 
\quad 
\bar{G}=G(\xi^1,\xi^2, \partial_1S_2-\partial_1S_2).
\end{align}
We call this system of equations ``generalized HJ equations" 
in the (1/2)-formalism.

The emergence of $S_0$ can be understood from a 
 gauge symmetry of the Nambu equations of motion, 
which was stressed by Nambu himself, but has often been 
discarded by later workers. 
The system of the equations \eqref{nambueq} is invariant under 
transformations $(H, G)\rightarrow (H',G')$ of the two Hamilton functions 
such that
\begin{align}
H\delta G-H'\delta G'=\delta \Lambda,
\label{Ngauge1}
\end{align}
where a generating function $\Lambda$ is an arbitrary function of 
$G$ and $G'$, satisfying
\begin{align}
\frac{\partial \Lambda}{\partial G}=H, 
\quad 
\frac{\partial \Lambda}{\partial G'}=-H'. 
\label{Ngauge2}
\end{align}
This property ensures that $\frac{\partial(H',G')}{\partial (H,G)}=1$ or equivalently, in terms of the partial derivatives 
with respect to $\xi^i$, 
\begin{align}
\partial_iH\partial_jG-\partial_jH\partial_iG=\partial_iH'\partial_jG'-
\partial_jH'\partial_iG', 
\label{Ngauge3}
\end{align}
and consequently the r.h.sides of \eqref{nambueq} 
are invariant under $(H,G)\rightarrow (H',G')$. Now, if the pair $(H, G)$ is 
replaced by $(H',G')$, the r.h.side of \eqref{1/2-HJ} is equal to
\[
\bar{H}'\partial_i\bar{G}'+\partial_iS_0=\bar{H}\partial_i\bar{G}-
\partial_i\Lambda+\partial_iS_0. 
\]
Therefore, the generalized HJ equations are invariant if we simultaneously 
shift $S_0$ by $S_0\rightarrow S_0'=S_0+\Lambda$. 
It should be kept in mind here that, though this gauge 
symmetry is analogous to that of electromagnetism, 
the degree of gauge freedom is weaker than the latter. 
The reason is that the form $H\partial_iG$ is {\it not} 
sufficiently general for representing an arbitrary given 
vector field in the form $A_i=H\partial_iG$: in order to exhaust the whole range 
of a vector field in this way, 
we would have to introduce multiple pairs of $(H_a,G_a)$ by 
extending to $A_i=\sum_aH_a\partial_iG_a$. 
But this weaker gauge symmetry is sufficient, at least,  
for the purpose of ensuring the symmetrical roles of 
two Hamiltonians in Nambu mechanics.

In addition to this, the system of generalized 
HJ equations itself has 
a gauge symmetry of its own with fixed $(H,G)$, under ($\mu=1,2,0$, $\partial_t=\partial_0$)
\begin{align}
S_{\mu}\rightarrow S_{\mu}+\partial_{\mu}\lambda
\end{align}
with an arbitrary scalar function $\lambda=\lambda(\xi_1,\xi_2,t)$. 
The above shift of $S_0$ associated with 
the gauge transformation of $(H, G)$ can 
be compensated by the second 
gauge transformation with $
\partial_t\lambda=-\Lambda$, which in turn 
induces the shift $\partial_i\lambda$ of the spatial components $S_i$. In order to distinguish these 
two gauge symmetries, we call the first one ``\,N\,"-gauge symmetry, and 
the second one ``\,S\,"-gauge symmetry. 

Now, we emphasize that one of prerequisites for step III is 
that, if $\bigl(\xi^1(t), \xi^2(t)\bigr)$ are chosen to be 
the solutions for the Nambu equations 
of motion as functions of time and are substituted into 
the generalized HJ equations, 
$\xi^3$ also becomes a function of time {\it and} automatically satisfies the Nambu equation as a consequence of \eqref{1/2-HJ}.  As in the case of ordinary HJ formalism, evidently, 
this is guaranteed 
in our case, because its derivation is attained 
merely by tracing back the above procedure conversely 
from the generalized HJ equations via 
the EE equations to the starting equation \eqref{xi3eq}.

Let us recast the generalized HJ equations in terms of differential forms. We have a natural 1-form on the 
base space $(\xi^1,\xi^2,t)$, 
\begin{align}
\bar{\Omega}^{(1)}\equiv S_id\xi^i+S_0dt\equiv S_{\mu}d\xi^{\mu}.
\end{align}
Then, the generalized HJ equations are expressed by the following 
equality on taking its exterior derivative:
\begin{align}
\bar{\Omega}^{(2)}\equiv d\bar{\Omega}^{(1)}&=(\partial_1S_2-\partial_2S_1)d\xi^1\wedge d\xi^2+(\partial_iS_0-\partial_0S_i)
d\xi^i\wedge dt\nonumber \\
&=\xi^3d\xi^1\wedge d\xi^2-\bar{H}(\partial_1\bar{G}d\xi^1+\partial_2\bar{G}d\xi^2)\wedge dt. 
\end{align}
The EE equation \eqref{1/2-EE} is equivalent to the vanishing condition for the 3-form $\bar{\Omega}^{(3)}=d\bar{\Omega}^{(2)}$:
\begin{align}
0=\bar{\Omega}^{(3)}=\partial_t\xi^3 d\xi^1\wedge d\xi^2 \wedge dt
-(\partial_1\bar{H}\partial_2\bar{G}-\partial_2\bar{H}\partial_1\bar{G})
d\xi^1\wedge d\xi^2 \wedge dt. 
\end{align}
It is to be noted that, unlike ordinary Hamilton mechanics, 
the fiber here is not directly related to tangent planes of the 
base space: the phase space 
in the present $(1/2)$-formalism may rather be called a 
``vorticity bundle" 
 instead of a cotangent bundle 
in the usual case. 

On the other hand, 
since $\partial_i\bar{G}=\partial_iG+\partial_i\xi^3\partial_3G$ and $d\xi^3=\partial_1\xi^3d\xi^1+\partial_2\xi^3d\xi^2$, 
the 2-form $\bar{\Omega}^{(2)}$ can be regarded as the projection, 
 from the three-dimensional phase space to 
the two-dimensional base space, each adjoined with time,  
of the following 2-form on the phase space $(\xi^1,\xi^2,\xi^3,t)$:
\begin{align}
\Omega^{(2)}\equiv \xi^3d\xi^1\wedge d\xi^2-HdG\wedge dt,
\label{Omega2}
\end{align}
where $dG$ is regarded as a differential 1-form on the 
phase space treating all of $\xi^1,\xi^2,\xi^3$ and $t$ as 
independent variables. This coincides with the 2-form that was 
used for defining the action integral (Ref. \cite{takh}). Then
 the closed (and exact) 3-form\footnote{The closed 3-form $\Omega^{(3)}$ 
 was first considered in \cite{esta}. Note that the 
 apparent emergence of the world ``surface"  $\xi^i(t, s_1)$  as mentioned in the previous section 
arises when we construct an integral invariant (Ref. \cite{takh} ) corresponding 
to \eqref{Omega2}. }
 $d\Omega^{(2)}$ \begin{align}
\Omega^{(3)}\equiv d\Omega^{(2)}= d\xi^1\wedge d\xi^2 
\wedge d\xi^3-dH\wedge dG\wedge dt
\end{align}
has a null field  
associated with the Nambu equations of motion: 
\begin{align}
\tilde{X}=\sum_{i=1}^3X^i\partial_i+\frac{\partial}{\partial t}: 
\quad i_{\tilde{X}}(\Omega^{(3)})=0.
\end{align}
Thus, up to step II of the present (1/2)-formalism, the structures and 
the relation between the 
EE equation and the HJ equations are almost 
parallel to the case of ordinary Hamilton 
mechanics, if 
the orders of the corresponding differential forms are 
increased by 1: $(\bar{\omega}^{(1)}, S) \rightarrow (\bar{\Omega}^{(2)}, \bar{\Omega}^{(1)}), 
(\omega^{(2)},\omega^{(1)})\rightarrow 
(\Omega^{(3)}, \Omega^{(2)})$.

\subsection{A difficulty:  From infinitesimal to finite canonical transformations}
At this juncture, let us 
 examine the property of $\Omega^{(2)}$ in 
three-dimensional phase space under a 
general time-dependent canonical transformation,  
a problem that,  to the author's knowledge, has not been 
studied in the literature. 
Our purpose is to 
consider whether we can expect that the similarities mentioned at the end of the previous subsection continue to the next step III. 
The general form of the infinitesimal 
canonical transformation for canonical coordinates is 
\begin{align}
\delta\xi^i=\{K, L, \xi^i\}, 
\end{align}
where $K$ and $L$ are arbitrary time-dependent functions. 
By a straightforward calculation, we find
\begin{align}
&\delta\bigl(\xi^3d\xi^1\wedge d\xi^2)=d\Sigma-\partial_tLdK\wedge dt
+\partial_tKdL\wedge dt,\\
&\Sigma\equiv KdL+\xi^3\Bigl(\frac{\partial(K,L)}{\partial(\xi^2,\xi^3)}d\xi^2
-\frac{\partial(K,L)}{\partial(\xi^3,\xi^1)}d\xi^1\Bigr).
\end{align}
This implies that the variation $\delta \Omega^{(2)}$ 
can be exact such that $\delta(\Omega^{(3)})=0$:
\begin{align}
\delta (\xi^3d\xi^1\wedge d\xi^2-HdG\wedge dt)
=d\Sigma, 
\nonumber
\end{align}
 {\it provided} that we choose, say,  
$L=G$ and
$
\delta H=\partial_tK, \delta G=0\nonumber 
$, 
which lead to
\[
-\delta (HdG)\wedge dt-\partial_tLdK\wedge dt+\partial_tKdL\wedge dt
=-(\delta H)dG\wedge dt+\partial_tKdG\wedge dt=0. 
\]
In other words, the infinitesimal time development 
described by the Nambu equations can be unfolded to 
initial time, as in the case of the ordinary 
Hamilton equations of motion. This can also be checked directly by performing 
the same canonical transformation on the Nambu equations themselves. 

Since a finite-time evolution can be generated by 
successive infinitesimal developments, we are justified in assuming that 
there exists a finite canonical transformation that 
unfolds the time development by sending $H$ to zero, with some 
finite 1-form $\Sigma'$ satisfying
\begin{align}
\xi^3d\xi^1\wedge d\xi^2-\bar{H}d\bar{G}\wedge dt=d\Sigma'
+Q_3dQ_1\wedge dQ_2, 
\end{align}
where $(Q_1, Q_2,Q_3)$ are appropriate canonical 
variables corresponding to the initial conditions. Note that here we have to 
regard $\xi^3$ as being projected to the space $(\xi^1,\xi^2,t)$, 
since we are implicitly assuming that 
conditions are appropriately given such that the trajectory 
is uniquely determined. This is analogous to the 
corresponding property of ordinary Hamilton mechanics
\[
p_idq^i-\bar{H}dt=dS+P_idQ^i, 
\]
which implies that we must regard the generating function 
$S=S(q,Q)$ as a function of 2$n$ 
independent variables $q^i$ and $Q^i$. This is legitimate, 
since there are 2$n$ independent canonical variables. Hence for a given 
$S$ satisfying the HJ equation $\partial_tS=-\bar{H}$,  
we obtain the standard relations that characterize the unfolding canonical transformation 
in concrete form: 
\[
\frac{\partial S}{\partial q^i}=p_i, \quad 
\frac{\partial S}{\partial Q^i}=-P_i, 
\] 
the latter of which is nothing but 
the Jacobi condition,  Eq. \eqref{jacobi}. 

Now 
going back to the case of Nambu mechanics, 
we notice a critical difference that there are only three independent 
canonical variables. Therefore we cannot treat both $dQ_1$ and $dQ_2$ as differentials that are independent of $d\xi^{\mu}$ $(\mu=1,2,0)$. Thus, one, say $Q_2$, out of  the pair $(Q_1,Q_2)$,  must be 
regarded as a dependent variable, namely, as a function 
of $(\xi^1,\xi^2, t)$, involving 
one constant parameter that we denote by $P$. Note that the last 
parameter is still necessary for representing the number (=3) of degrees of freedom 
for initial conditions for determining the trajectory 
uniquely as a function of time.  Then, by defining
\begin{align}
\Sigma'=&\Sigma'_{\mu}d\xi^{\mu}+\Sigma'_{Q_1}dQ_1, 
\nonumber \\
d\Sigma'=&(\partial_1\Sigma'_2-\partial_2\Sigma'_1)d\xi^1
\wedge d\xi^2-(\partial_t\Sigma'_i-\partial_i\Sigma'_0)
d\xi^i\wedge dt\nonumber \\
&+
\Bigl(\partial_i\Sigma'_{Q_1}-\frac{\partial \Sigma'_i}{\partial Q_1}\Bigr)
d\xi^i\wedge dQ_1+\Bigl(\partial_t\Sigma'_{Q_1}-\frac{
\partial \Sigma'_0}{\partial Q_1}\Bigr)dt\wedge dQ_1\nonumber 
\end{align}
we obtain, formally,
 \begin{align}
&\partial_1\Sigma'_2-\partial_2\Sigma'_1=\xi^3,
\label{Sigma'0} \\
&\partial_t\Sigma'_i-\partial_i\Sigma'_0=\bar{H}\partial_i\bar{G}, 
\label{Sigma'1}\\
&\partial_i \Sigma'_{Q_1}-\frac{\partial \Sigma'_i}{\partial Q_1}=Q_3\partial_iQ_2, 
\quad \partial_t\Sigma'_{Q_1}-\frac{\partial\Sigma'_0}{\partial Q_1}=Q_3\partial_tQ_2.
\label{Sigma'2}
\end{align}
Comparing Eqs. \eqref{Sigma'0} and \eqref{Sigma'1} with the generalized HJ equations \eqref{1/2-HJ0} and 
\eqref{1/2-HJ} previously obtained, respectively, 
we can identify that $\Sigma'_{\mu}=S_{\mu}$ up to 
gauge degrees of freedom. On the other hand, 
Eq. \eqref{Sigma'2} specifies the dependence on the initial 
conditions, giving an implicit characterization of the 
required finite canonical transformation. 
Therefore these two relations replace 
the Jacobi conditions of the ordinary case.  Using the S-gauge symmetry that is now extended to $\Sigma'_{Q_1}\rightarrow 
\Sigma'_{Q_1}+\partial\lambda/\partial Q_1$, we can 
set $\Sigma'_{Q_1}=0$ without losing generality. 
Then the residual gauge symmetry is the previous 
S-gauge symmetry. By a further redefinition 
$\Sigma'_{\mu}=S_{\mu}
\rightarrow S_{\mu}
-Q_1Q_3\partial_{\mu}Q_2(\xi^1,\xi^2,t)$, 
the r.h.sides  
of \eqref{Sigma'2} can be sent to zero.  
This is analogous to our remarks on the condition 
\eqref{jacobi} 
in the previous section, concerning the 
shift $S\rightarrow S+P_iQ^i$ of the action function 
in the ordinary case.  However, the situation here is quite different 
in the sense that $Q_2$ is in general an unknown function. This exhibits 
a difficulty in characterizing finite 
canonical transformations in Nambu mechanics. 

A possible way out is to choose $Q_2$ from the beginning to 
be a known function. Since we have two conserved fields 
$(\bar{H}, \bar{G})$, a natural choice would be $Q_2=\bar{G}$ 
in view of our choice $\delta G=0$ for the infinitesimal transformation discussed at the beginning of this subsection. 
In this case, the additional constant $P$ should be 
taken as the numerical value of $\bar{G}$ itself. 
However, a difficulty remains. Remember that 
$\bar{G}$ depends, in general, on $\xi^3=\partial_1S_2-\partial_2S_1$. This means that the r.h.sides of 
the first relation in Eq. \eqref{Sigma'2} involve, in general, 
second derivatives of $S_i$. 
But the generalized HJ equation \eqref{Sigma'1} ($\Sigma'_{\mu}=S_{\mu}$) does not 
directly dictate their property: to use them for 
deriving the Nambu equation, we are led to perform partial derivatives on both sides. Then the r.h.sides of \eqref{Sigma'1} involve higher derivatives than second. 
Thus we are led to performing 
another partial derivative on both sides, thereby 
inducing yet higher derivatives, ad infinitum. 
In contrast to this, \eqref{jacobi} in the ordinary case does not 
involve derivatives $\partial_iS$ at all.

{\it Conclusion}: The relations 
\eqref{Sigma'2} are still too implicit to characterize finite 
canonical transformations for the purpose of regaining the equations of motion, unfortunately, in completing the Jacobi procedure 
in general. 

\vspace{0.4cm}
\subsection{Step III}
Consideration of the previous subsection suggests that we should 
look at solving the generalized HJ equation in order 
to obtain explicit results more 
suitable to step III than \eqref{Sigma'2} regarding 
integration constants. We will present 
two approaches. The first one assumes that 
by suitable N-gauge and/or canonical transformation, 
$G$ is chosen to be independent of $\xi^3$: 
$\partial_3G=0$. In principle, this requirement 
is not restrictive, but practical applicability 
may be confined to relatively simpler theories. 
The second approach applies to the general case without any 
restriction at the beginning, but requires us, instead, to 
make a further fiber-decomposition of 
the two-dimensional base space to 
(1/1). 

\vspace{0.2cm}
\noindent
{\bf Approach (i)}: $\partial_3G=0$ case

A drastic simplification occurs when $G$ is independent of 
$\xi^3$: $\partial_3G=0$, or equivalently $\bar{G}=G$ identically, 
which implies that $\partial_t\bar{G}=0$ since 
the whole dependence on time arises through $\xi^3$.
Let us choose the gauge $S_0=0$ using the S-gauge 
symmetry of the generalized HJ equations. 
Then, if we set
\begin{align}
S_i=-S\partial_iG +g_i, 
\end{align}
by introducing a function $S=S(\xi^i, t)$ that satisfies
\begin{align}
\partial_tS=-\bar{H}, 
\label{G3HJ}
\end{align}
Eq. \eqref{1/2-HJ} reduce to 
\[
\partial_tg_i=0, 
\]
and 
we have 
$
\xi^3=\partial_1G\partial_2S-\partial_2G\partial_1S+\partial_1g_2-\partial_2g_1. 
$
Using the residual (time-independent) S-gauge symmetry, furthermore,
 we can set 
$g_i=\epsilon^{ij}\partial_jg$ in terms of a time-independent 
scalar function $g$: $\partial_1g_2-\partial_2g_1=-\partial_i\partial^ig$. 
Then, $g_i$ can be absorbed into $S$ by making a 
redefinition
\[
S\rightarrow S+F
\]
where $F$ is defined such that $F\partial_iG=g_i=\epsilon^{ij}\partial_jg$, which is solved as 
$d_GF=\partial^2g$ 
with
\begin{align}
d_G\equiv \partial_1G\partial_2-
\partial_2G\partial_1, 
 \end{align} 
assuming that at least 
one $\partial_iG$ is not zero. Therefore we can actually set $g_i=0$ without 
losing generality, and obtain
\begin{align}
\xi^3=d_GS.
\label{Gdiff}
\end{align}
We have arrived almost at the usual HJ equation: the only difference 
is that, instead of the projected generalized momenta as just plain 
partial derivatives of $S$, we now have Eq. \eqref{Gdiff} corresponding to
an infinitesimal shift $\delta \xi^i=-\epsilon^{ij}\partial_jG$ 
in the base (configuration) space, which preserves  
$G$: $d_GG=0$, identically. 

We can now complete step III for this system. Suppose we have a 
solution to the generalized HJ equation \eqref{G3HJ}, supplemented by Eq. \eqref{Gdiff}, which has one arbitrary integration constant $Q_1$. 
Then in analogy with the case of ordinary HJ equations, 
we impose 
\begin{align}
\frac{\partial S}{\partial Q_1}=Q_3, 
\label{1/2-jacobi}
\end{align}
by introducing an additional constant $Q_3$. If we further require that 
$G$ is preserved and hence impose $G=\bar{G}=Q_2$ with 
$Q_2$ being an additional 
time-independent constant, we can solve these two 
conditions to obtain $\xi^i$ as functions, $\xi^i(t; Q_1, Q_2, Q_3)$, of time 
with three parameters. 
Solvability is ensured by requiring that \eqref{1/2-jacobi} is not 
invariant under the action of $d_G$, namely, 
\begin{align}
\frac{\partial^2S}{\partial \xi^1\partial Q_1}\partial_2G-
\frac{\partial^2S}{\partial \xi^2\partial Q_1}\partial_1G\ne 0, 
\label{solveIII}
\end{align}
for it to be independent of the condition $G=Q_2$. 
Condition \eqref{1/2-jacobi} can also 
be derived from consideration of the canonical 
transformation of the previous subsection, due to the 
simplification that now 
$G=Q_2$ does not 
depend on $\xi^3$, since then 
$
\frac{\partial S_i}{\partial Q_1}=-\partial_iG\frac{\partial S}{\partial Q_1} 
$
and consequently the first of \eqref{Sigma'2}, with $\Sigma'_{Q_1}=0$, reduces to \eqref{1/2-jacobi}, while the second becomes trivial in the present 
gauge $S_0\, (=\Sigma'_0)=0$. 

Then we can derive the Nambu equations of motion by extending 
the method that is explained in the ordinary case: 
by taking a (total) time derivative of \eqref{1/2-jacobi} and $G=Q_2$, we 
obtain
\begin{align}
&\frac{\partial^2S}{\partial Q_1\partial t}+
\frac{\partial^2S}{\partial\xi^1\partial Q_1}
\frac{d\xi^1}{dt}+\frac{\partial^2S}{\partial\xi^2\partial Q_1}
\frac{d\xi^2}{dt}=0, \label{III1} \\
&\partial_1G\frac{d\xi^1}{dt}+\partial_2G\frac{d\xi^2}{dt}=0. 
\label{III2}
\end{align}
On the other hand, since \eqref{G3HJ} is satisfied for an 
arbitrary constant $Q_1$ by its definition, we also have
\[
\frac{\partial^2S}{\partial Q_1\partial t}=
-\frac{\partial H}{\partial \xi^3}
\Bigl(\partial_1G\frac{\partial^2S}{\partial \xi^2
\partial Q_1}-\partial_2G
\frac{\partial^2S}{\partial\xi^1\partial Q_1}\Bigr), 
\]
remembering that the dependence on $Q_1$ of $\bar{H}$ resides 
only in $\xi^3$. 
Using this result and \eqref{III2}, 
\eqref{III1} reduces, due to condition \eqref{solveIII}, to
\begin{align}
&\frac{d\xi^1}{dt}=-\partial_3H\partial_2G, \\
&\frac{d\xi^2}{dt}=\partial_3H\partial_1G, 
\end{align}
which are nothing but the Nambu equations under the condition $\partial_3G=0$.

It remains to see to what extent the condition $\partial_3G=0$ 
is attainable. Consider first the possibility of an N-gauge 
transformation alone. Since 
the general N-gauge transformation $(H,G)\rightarrow 
(H',G')$ must satisfy \eqref{Ngauge3}, the necessary and 
sufficient condition for the ``axial gauge" $\partial_3G'=0$ can be 
stated as 
\[
\partial_3H\partial_1G-\partial_1H\partial_3G=\partial_3H'\partial_1G', 
\quad 
\partial_3H\partial_2G-\partial_2H\partial_3G=\partial_3H'\partial_2G'
\]
for some $H'$. This shows that the ratio 
\[
R\equiv \frac{\partial_3H\partial_1G-\partial_1H\partial_3G}
{\partial_3H\partial_2G-\partial_2H\partial_3G}\]
must be independent of $\xi^3$. Obviously, 
there are an infinite number of possibilities for $(H,G)$ for which this is not satisfied. Note that this exemplifies the weakness of 
the N-gauge symmetry compared with the case 
of electromagnetism. 
Thus, from the viewpoint of 
the N-gauge transformation, the attainability is restricted to a special 
class. For example, the case of the Euler top belongs to this class: 
$R
=\bigl(\xi^1(1-I_1/I_3)\bigr)/\bigl(\xi^2(1-I_1/I_2)\bigr)$. 
Indeed by an N-gauge transformation with 
the generator $\Lambda=H^2/2I_3$, giving 
$
(H',G')=(H,G-H/I_3)$, 
we have $\partial_3G'=0$. 

On the other hand, it is evident, in principle, that 
we can use a finite
canonical (or volume-preserving) coordinate transformation $(\xi^1,\xi^2,\xi^3)\rightarrow (\xi'^1, \xi'^2, \xi'^3)$ 
in order to bring $G$ to $G'$ such that $\partial_{3'}G'=0$, at least, 
locally. 
If a concrete form of $G$ is given, 
we will be able, in general, to combine N-gauge and canonical 
coordinate transformations in bringing the system to 
fit our requirement, unless it is too complicated.


\vspace{0.2cm}
\noindent
{\bf Approach (ii)}: (1/1/1)-formalism

There is another possibility for simplification without 
any special requirement for $H, G$. In ordinary 
HJ theory, we can reduce the HJ equation to a 
time-independent one by setting the Hamiltonian to be 
a constant $H=E$ at the beginning; the HJ equation then reduces to 
$H(\partial_q\tilde{S}, q)=E$ for a reduced function $\tilde{S}(q)$ 
that is related to $S$ 
by $S(q,t)=-Et+\tilde{S}(q)$ where only the first term 
involves time. By choosing $E$ as one 
of the integration constants, the Jacobi conditions in step III 
necessary involve the condition 
\begin{align}
t-t_0=\frac{\partial \tilde{S}}{\partial E}
\label{tcond-ord}
\end{align}
from the Jacobi condition $\partial S/\partial E=t_0$ with 
$t_0$ being one of the additional constants. Together with other 
Jacobi conditions associated with the other integration constants, 
\eqref{tcond-ord} determines the 
trajectories as functions of time, satisfying the equations of motion.

In the present case, by setting $H=E$, the generalized HJ equations 
reduce to
\begin{align}
\partial_tS_i-\partial_iS_0=E\partial_iG
\end{align}
which are trivially solved as
\begin{align}
S_0=-EG+f(t), \quad 
S_i=g_i
\end{align}
where $f$ is an arbitrary function of time and the 
$g_i$ are time-independent functions. In terms of the latter,  
$\xi^3$ is given by
\begin{align}
\xi^3=\partial_1g_2-\partial_2g_1
\end{align}
which is constrained by
\begin{align}
\bar{H}=H\bigl(\xi^1,\xi^2,\xi^3(\xi^1,\xi^2; E)\bigr)=E.
\label{Econd}
\end{align}
Obviously, the S-gauge symmetry actually 
allows us to set $f=0$ without losing 
generality. Therefore in comparison with the case of ordinary 
time-independent HJ equations,  there is a crucial difference that 
time is now {\it completely} eliminated from the scene. Of course, it is still guaranteed 
that if $\xi^1(t)$ and $\xi^2(t)$ are given 
as functions of time satisfying 
the Nambu equations of motion, $\xi^3$ determined through these 
relations 
automatically obeys the Nambu equation 
\begin{align}
\frac{d\xi^3}{dt}=\partial_1H\partial_2G-\partial_2H\partial_1G. 
\nonumber 
\end{align}

In this situation, we introduce further decomposition 
of the base space $(\xi^1,\xi^2)$ into (1/1) fiber bundle in which 
$\xi^2$ is  
regarded as a one-dimensional fiber and $\xi^1$ as a base space, 
respectively, in order to regain the 
time $t$ on the scene. 
In other words, the two-dimensional base space $(\xi^1,\xi^2,t)$ 
regarded as the configuration space up to this point is now treated as the phase space. 
Then by making the projection $\xi^2=\xi^2(\xi^1,t)$ we can introduce 
a further HJ structure, involving time, such that the Nambu equations of motion 
are obtained by the Jacobi-like prescription. 
What we do is expressed symbolically by 
$(1/2)\rightarrow (1/1/1)$. This procedure 
itself should be regarded as a part 
of step III in establishing prescriptions to obtain 
the equations of motion from our generalized HJ equations. 

For this purpose, 
we must have an appropriate 1-form, which we denote 
by $\Omega'^{(1)}$, 
 on 
$(\xi^1,t)$ such that the requirement 
$\Omega'^{(1)}=dT$ with some 0-form $T$ gives 
an HJ-like equation in the base space $(\xi^1,t)$. 
Now the EE equation in the present problem is obtained 
by the same method as before 
\begin{align}
\partial_t\xi^2=\partial_3H\partial_1\bar{\bar{G}}, 
\label{EExi-2}
\end{align}
by rewriting the Nambu equation of motion 
\begin{align}
\frac{d\xi^2}{dt}=\partial_3H\partial_1G-\partial_1H\partial_3G\nonumber 
\end{align}
a a field equation for $\xi^2(\xi^1,t)$ using 
\begin{align}
\frac{d\xi^1}{dt}=\partial_2H\partial_3G-\partial_3H\partial_2G.\nonumber 
\end{align}
Here and in what follows, double bars over symbols mean,  
e.g.,
\begin{align}
\bar{\bar{G}}(\xi^1,t)=\bar{G}\bigl(\xi^1,\xi^2(\xi^1,t)\bigr)
=G\Bigl(\xi^1,\xi^2(\xi^1,t), \xi^3\bigl(\xi^1,\xi^2(\xi^1,t)\bigr)\Bigr).
\label{doublebar}
\end{align}

\noindent
[proof of \eqref{EExi-2}]:  We note that for $i=1$ and $i=2$, because of constraint \eqref{Econd}, 
\begin{align}
0=\partial_i\bar{H}=\partial_iH+\partial_3H\partial_i\xi^3.
\end{align}
and also 
\begin{align}
1=\partial_3H\frac{\partial\xi^3}{\partial E}.
\end{align}
Then,  we have
\begin{align}
\partial_i\bar{G}=\partial_iG+\partial_3G\partial_i\xi^3=\frac{1}{\partial_3H}(\partial_3H\partial_iG-\partial_iH\partial_3G).
\label{partial-bar}
\end{align}
Using this result, we rewrite the equation of motion for $\xi^2$ with a further projection $\xi^2=\xi^2(\xi^1,t)$, 
\begin{align}
\frac{d\xi^2}{dt}=\partial_t\xi^2+\partial_1\xi^2\frac{d\xi^1}{dt}=
\partial_t\xi^2-\partial_1\xi^2\partial_3H
\partial_2\bar{G}. 
\end{align}
On the other hand, this is also rewritten as
\begin{align}
\frac{d\xi^2}{dt}=
\partial_3H\partial_1\bar{G}. 
\end{align} Hence, 
\[
\partial_t\xi^2=\bigl[\partial_3H(\partial_1\bar{G}+\partial_1\xi^2
\partial_2\bar{G})\bigr]_{\xi^2=\xi^2(\xi^1,t)}=
\partial_3H\partial_1\bar{\bar{G}}.
\]
[Q.E.D.]

These results show that we can define the following closed 
2-form on $(\xi^1,\xi^2, t)$:
\begin{align}
\bar{\Omega}'^{(2)}
\equiv \frac{\partial\xi^3}{\partial E}d\xi^1\wedge d\xi^2-d\bar{G}\wedge dt=\frac{1}{\partial_3H}d\xi^1\wedge d\xi^2-d\bar{G}\wedge dt. \label{Omega'2}
\end{align}
Namely, we now consider a symplectic structure $(\partial\xi^3/\partial E)d\xi^2\wedge d\xi^2=(1/\partial_3H)d\xi^1\wedge d\xi^2$ in the phase space $(\xi^1,\xi^2)$. 
The EE equation is then nothing but the vanishing condition for 
$\bar{\Omega}'^{(2)}$ 
under the projection by $\xi^2=\xi^2(\xi^1,t)$ from $(\xi^1,\xi^2,t)$ 
to $(\xi^1,t)$:
\begin{align}
\bar{\bar{\Omega}}'^{(2)}\equiv \bar{\Omega}'^{(2)}|_{\xi^2=\xi^2(\xi^1,t)}=
\Bigl(\frac{1}{\partial_3H} \partial_t\xi^2-\partial_1\bar{\bar{G}}\Bigr)d\xi^1\wedge dt=0.
\end{align}
The desired 1-form $\Omega'^{(1)}$ is then obtained as
\begin{align}
&d\Omega'^{(1)}=\bar{\Omega}'^{(2)},\\
&\Omega'^{(1)}=p_1d\xi^1-\bar{G}dt, 
\end{align}
where 
\begin{align}
p_1=p_1(\xi^1,\xi^2)=-\int^{\xi^2}\frac{dx}{\partial_3H(\xi^1,x)}, 
\label{p1}
\end{align}
which satisfies $
\partial_tp_1=-\partial_t\xi^2/\partial_3H$. 
Thus we have a generalized HJ equation for the 0-form $T$ 
by requiring $dT=\Omega'^{(1)}$ under the projection $\xi^2=\xi^2(\xi^1,t)$: 
\begin{align}
&\partial_tT+\bar{\bar{G}}=0,\label{HJ-T}\\
&p_1=\partial_1T, \label{HJ-p}
\end{align}
the latter of which enables us to express $\xi^2$ in terms of $\partial_1T$ through \eqref{p1}. 
Thus what we have achieved is that the base space $(\xi^1,\xi^2)$ 
is interpreted as a quasi-cotangent bundle on the 
one-dimensional base space $\xi^1$ equipped 
with fibers represented by 
$p_1$. In this way, we have recovered time on the scene. 

We can now finish step III in the present case. Suppose we have 
a complete solution of \eqref{HJ-T} with one integration 
constant $Q_1$. Then in order to extract the solution $\xi^1=
\xi^1(t; Q_1, P, E)$ with three integration constants to 
the equation of motion, we impose a Jacobi-type condition 
\begin{align}
\frac{\partial T}{\partial Q_1}=P
\end{align}
by introducing one additional constant $P$. 
The solubility of $\xi^1$ from this condition 
\[
\frac{\partial T}{\partial Q_1\partial \xi^1}\ne 0
\]
must be assumed here. 
Note also that 
the constant $E$ is inherited already from the 
constraint \eqref{Econd}. 
Then, using Eqs. \eqref{HJ-T} and \eqref{HJ-p}, we have
\begin{align}
&0=\frac{d}{dt}\frac{\partial T}{\partial Q_1}=
\frac{\partial^2T}{\partial Q_1\partial t}+
\frac{\partial^2T}{\partial Q_1\partial \xi^1}
\frac{d\xi^1}{dt}=-\frac{\partial \bar{\bar{G}}}{\partial Q_1}
+\frac{\partial^2T}{\partial Q_1\partial \xi^1}\frac{d\xi^1}{dt}, \nonumber \\
&\frac{\partial\bar{\bar{G}}}{\partial Q_1}=
\partial_2\bar{G}\frac{\partial \xi^2}{\partial Q_1}=\partial_2\bar{G}
\frac{\partial \xi^2}{\partial p_1}\frac{\partial p_1}{\partial Q_1}=-
\frac{\partial^2T}{\partial Q_1\partial \xi^1}\partial_2\bar{G}\partial_3H, \nonumber
\end{align}
where $\partial_2\bar{G}$ should be understood as 
being substituted 
$\xi^2=\xi^2(\xi^1,t)$ after performing the designated 
partial differentiation.
Thus, 
\begin{align}
\frac{d\xi^1}{dt}=-\partial_3H\partial_2\bar{G}=-(\partial_3H\partial_2G
-\partial_2H\partial_3G)
\end{align}
due to the relation \eqref{partial-bar}. As before, the 
equation for $\xi^2$ is a consequence of the EE equation 
\eqref{EExi-2} for $\xi^2$, and the equation for $\xi^3$ is a consequence 
of the generalized HJ equation obtained in step II. 
The last procedure, in the present case, 
is essentially equivalent to using the constraint \eqref{Econd} directly 
through $dH/dt=0$:
\[
\frac{d\xi^3}{dt}=-\frac{1}{\partial_3H}\Bigl(
\partial_1H\frac{d\xi^1}{dt}+\partial_2H\frac{d\xi^2}{dt}\Bigr)
=\partial_1H\partial_2G-\partial_2H\partial_1G. 
\]

{\it Remark}: Let us recall the original 
2-form $\bar{\Omega}^{(2)}$ 
corresponding to the EE equation in the previous step II, 
which reads under the constraint \eqref{Econd} 
\[
\bar{\Omega}^{(2)}=\xi^3d\xi^1\wedge d\xi^2-\bar{H}d\bar{G}\wedge dt
=E\Bigl(\frac{\xi^3}{E}d\xi^1\wedge d\xi^2-d\bar{G}\wedge dt\Bigr). 
\]
Thus $\bar{\Omega}'^{(2)}$, \eqref{Omega'2}, is obtained from $\bar{\Omega}^{(2)}/E$ 
by substituting $\partial\xi^3/\partial E$ in place of
 $\xi^3/E$. 
There is also a similar relation for 1-forms. In the present case, 
$
\bar{\Omega}^{(1)}$ takes the form
\[
\bar{\Omega}^{(1)}=g_1d\xi^1-E\bar{G}dt=E\Bigl(\frac{g_1}{E}d\xi^1-
\bar{G}dt\Bigr)
\]
where we have set $g_2=0$ by using the S-gauge symmetry 
without losing generality. 
In this gauge, we have 
\[
g_1(\xi^1,\xi^2)=-\int^{\xi^2} \hspace{-0.3cm} dx\, \xi^3(\xi^1,x), 
\]
which apparently corresponds to \eqref{p1}. 
Thus, again, $\Omega'^{(1)}$ is obtained from $\bar{\Omega}^{(1)}/E$ 
by the same replacement
$
\frac{\xi^3}{E}\rightarrow \frac{\partial\xi^3}{\partial E}. 
$
We interpret these relations as part of the analogue, for the case of the 
integration constant $E$,  of the prescription \eqref{tcond-ord} 
for restricting complete solutions of the ordinary 
HJ equations. 
It would be interesting to clarify more about these 
correspondences in geometrical terms. 

\section{(2/1)-formalism}
\setcounter{equation}{0}
We turn to the second possibility of decomposing phase space: one-dimensional base space with 
the coordinate $\xi^1$ and two-dimensional fibers 
described by the two-component fields $(\xi^2(\xi^1,t), \xi^3(\xi^1,t))$.  
Throughout this section, the convention for denoting 
independent variables is
\begin{align}
\hat{H}(\xi^1,t)\equiv 
H\bigl(\xi^1, \xi^2(\xi^1,t), \xi^3(\xi^1,t)\bigr), \nonumber 
\end{align}
 so that 
\[
\partial_1\hat{H}=\partial_1H+\partial_2H\partial_1\xi^2+
\partial_3H\partial_1\xi^3, 
\]
where (and in what follows unless otherwise stated) 
the substitution of $\xi^2=\xi^2(\xi^1,t)$ and 
$\xi^3=\xi^3(\xi^1,t)$ should be understood on the 
r.h.side 
after performing partial differentiations  
on functions without the hat symbol. 

\subsection{Step I}
By following the method explained in Sect. 2, we 
derive the EE equations for the field $\bigl(\xi^2(\xi^1,t), \xi^3(\xi^1,t)\bigr)$:
\begin{align}
&\partial_t\xi^2=\partial_3H\partial_1\hat{G}-\partial_3G\partial_1\hat{H},
\label{2/1-EE2}\\
&\partial_t\xi^3=\partial_1\hat{H}\partial_2G-
\partial_1\hat{G}\partial_2H. 
\label{2/1-EE3}
\end{align} 
It is guaranteed that 
if we substitute a 
solution $\xi^1=\xi^1(t)$ of the equation of motion 
ino the EE equations, they automatically 
give the equations of motion for $\xi^2$ and $\xi^3$. 
For example,
\begin{align}
\frac{d\xi^2}{dt}&=\partial_t\xi^2+
\partial_1\xi^2\frac{d\xi^1}{dt}=\partial_3H\partial_1\hat{G}-\partial_3G\partial_1\hat{H}+\partial_1\xi^2(\partial_2H\partial_3G-\partial_3H\partial_2G)\nonumber \\
&=\partial_3H\partial_1G-\partial_3G\partial_1H. 
\nonumber 
\end{align}
Of course, as we have already emphasized in Sect. 2, 
this merely traces back the derivation of the EE equations 
in reverse order. 

It is also easy to recast the EE equations in terms of 
differential forms. We first define two independent 
2-forms $(\Omega_2^{(2)}, \Omega_3^{(2)})$ in the original phase space $(\xi^1,\xi^2, \xi^3,t)$:
\begin{align}
\Omega_2^{(2)}&\equiv d\xi^2\wedge d\xi^1+
(\partial_3HdG-\partial_3GdH)\wedge dt \nonumber \\
&=d\xi^2\wedge d\xi^1+
(\partial_3H\partial_1G-\partial_3G\partial_1H)
d\xi^1\wedge dt+(\partial_3H\partial_2G-\partial_3G\partial_2H)
d\xi^2\wedge dt,\label{omega22}\\
\Omega_3^{(2)}&\equiv d\xi^3\wedge d\xi^1-
(\partial_2HdG-\partial_2GdH)\wedge dt \nonumber \\
&=d\xi^3\wedge d\xi^1-(\partial_2H\partial_1G-\partial_2G\partial_1H)d\xi^1\wedge dt
-(\partial_2H\partial_3G-\partial_2G\partial_3H)d\xi^3\wedge dt.
\label{omega23}
\end{align}
By making a projection to the base space $(\xi^1,t)$ 
with substitutions $\xi^2=\xi^2(\xi^1, t)$ and $ \xi^3=\xi^3(\xi^1,t)$,  
these 2-forms are given, respectively, as
\begin{align}
&\hat{\Omega}_2^{(2)}\equiv -\partial_t\xi^2d\xi^1\wedge dt
+(\partial_3H\partial_1\hat{G}-\partial_3G\partial_1\hat{H})d\xi^1\wedge dt, \\
&\hat{\Omega}_3^{(2)}\equiv -
\partial_t\xi^3d\xi^1\wedge dt-(\partial_2H\partial_1\hat{G}-
\partial_2G\partial_1\hat{H})d\xi^2\wedge dt. 
\end{align}
Thus the EE equations \eqref{2/1-EE2} and \eqref{2/1-EE3} 
coincide with the vanishing conditions for these projected 
2-forms. 

Furthermore, when $\xi^2$ and $\xi^3$ are treated 
as independent variables without projection, it is straightforward to confirm by explicit 
calculation that the vector differential operator 
$\tilde{X}$ defined in Sect. 3 satisfies the null condition both 
for these 2-forms simultaneously:
\begin{align}
i_{\tilde{X}}(\Omega_2^{(2)})=i_{\tilde{X}}(\Omega_3^{(2)})=0. 
\end{align}

These properties are almost parallel to those of the 
3-form $\Omega^{(3)}$ of the 
(1/2)-formalism, with decreased orders of the corresponding 
forms.  One difference is that, in the present case, 
vanishing (or null) conditions that lead to 
the EE equations are now characterized by 
two 2-forms instead of a single 3-form in the (1/2)-formalism. 
In the latter case, the line field corresponding to the 
vector $X^i$ is contained as the bounding edges of 
infinitesimal cubes associated with 
$\Omega^{(3)}$, while in the present (2/1) decomposition 
it is contained simultaneously in the boundaries of two infinitesimal squares 
associated with $(\Omega^{(2)}_2, \Omega^{(2)}_3)$. 
Another (and crucial) difference is that the 2-forms $\Omega^{(2)}_2$ and 
$\Omega^{(2)}_3$ are not closed, due to the presence 
of the second terms in their definitions. This shows that we cannot 
associate these 2-forms directly to variational principles in the 
three-dimensional (plus time) phase space $(\xi^1,\xi^2,\xi^3,t)$. 
It would be very interesting to clarify further the geometrical 
meaning of $(\Omega^{(2)}_2, \Omega^{(2)}_3)$. 

\subsection{Step II and III: Reduction to the {\rm (1/1/1)}-formalism}
Due to the fact that $\Omega^{(2)}_2$ and $\Omega^{(2)}_3$ are 
not closed, the next steps II and III must necessarily be modified, 
compared with the previous section,  in the 
present (2/1)-formalism. Since the base space is one-dimensional, 
it is natural to find connection with 
the (1/1/1)-formalism. 
Let us first examine whether the EE equations \eqref{2/1-EE2} and \eqref{2/1-EE3} themselves allow
$H$ (and $G$, if both necessary) as integration constants. 
We find that 
\begin{align}
&\partial_t\hat{H}
=\partial_2H\partial_t\xi^2+\partial_3H\partial_t\xi^3
=\partial_1\hat{H}(\partial_3H\partial_2G-\partial_2H\partial_3G),\\
&\partial_t\hat{G}
=\partial_2G\partial_t\xi^2+\partial_3G\partial_t\xi^3
=\partial_1\hat{G}(\partial_3H\partial_2G-\partial_2H\partial_3G).
\end{align}
Thus we can indeed choose $\hat{H}$ (and/or $\hat{G}$)   
as an integration constant separately for the partial differential 
equations \eqref{2/1-EE2} and \eqref{2/1-EE3}. 
In order to connect this system to the (1/1/1)-formalism we set only 
the first one, 
\begin{align}
\hat{H}=H\big(\xi^1,\xi^2(\xi^1,t),\xi^3(\xi^1,t)\big)=E
\end{align}
with constant $E$. Then the EE equations are rewritten as 
\begin{align}
&\partial_t\xi^2=\partial_3H\partial_1\hat{G}, \\
&\partial_t\xi^3=-\partial_2H\partial_1\hat{G}
\end{align}
using $\partial_1\hat{H}=\partial_1\hat{G}=0$. 
It is sufficient to consider the first one $\partial_t\xi^2$, 
noticing that it coincides with \eqref{EExi-2}, when 
$\bar{\bar{G}}$ is identified with $\hat{G}$ as it should be 
since the definition \eqref{doublebar} amounts, after all, to 
making $G$ a field on one and the same base space $(\xi^1,t)$. 
In connection with this, it is to be noted that the closed 2-form 
$\bar{\Omega}'^{(2)}$ of the (1/1/1) formalism is naturally obtained from \eqref{omega22} after the projection 
from $(\xi^1,\xi^2,\xi^3,t)$ to $(\xi^1,\xi^2,t)$ under the 
constraint $H=E$ as
\[
\bar{\Omega}'^{(2)}=-\frac{1}{\partial_3H}\Omega^{(2)}_2\Bigr|_{\xi^3=
\xi^3(\xi^1,\xi^2;E)}, 
\]
on using 
$
\partial_iH=-\partial_3H\partial_i\xi^3$ and $
\partial_i\bar{G}=\partial_iG+\partial_3G\partial_i\xi^3. 
$
We can thus repeat the same arguments as in the 
(1/1/1)-formalism, finishing steps II and III simultaneously. 
Thus the equation of motion for $\xi^1(t)$ is derived in exactly 
the same way: the existence of three independent constants 
$Q_1, P$ and $E$ ensures that we have general solutions 
for the Nambu equations of motion. 

Finally, for a better appreciation of the necessity of the 
(1/1/1)-formalism in the present context, we 
note the following. If we choose from the beginning 
both $\hat{H}$ and $\hat{G}$ simultaneously as integration 
constants for the EE equations, the latter reduces  
to $\partial_t\xi^2=0=\partial_t\xi^3$. Thus, time is apparently 
eliminated from the scene again. Although the origin is somewhat different from 
what happens in the time-independent solution in the case of the (1/2)-formalism, this motivates us to the (1/1/1)-formalism.

\section{Examples}
\setcounter{equation}{0}
We have established a generalized Hamilton-Jacobi theory 
for Nambu mechanics.  Two different formalisms were 
presented: the first one under the 
requirement $\partial_3G=0$ is called the (1/2)-formalism, and the second which 
is something analogous to the ordinary time-independent Hamilton-Jacobi theory 
is called the (1/1/1)-formalism; the case of the 
(2/1) decomposition was 
also reduced to the (1/1/1) formalism. We have discussed only the 
simplest case of three (+time)-dimensional 
phase space. 
The basic ideas, in principle, can be 
extended straightforwardly to Nambu mechanics of higher orders.\footnote{It is to be noted here that the extension to 
$3n$ ($n>1$) dimensions by introducing $n$ coupled 
triplets as originally suggested by Nambu is {\it not} feasible, for the  reason that such ``canonical" structures cannot 
be preserved by the equations of motion and 
canonical transformations defined by the corresponding brackets, 
as signified by the violation of the so-called fundamental identity. 
This situation is in marked contrast to that of the usual Hamilton mechanics. The negative comment given in Ref. \cite{esta} about the 
possibility of HJ formalism 
is also related to this difficulty. 
}  
The resultant formalisms, however, become increasingly 
complicated since we have various 
possibilities of decomposing phase spaces into 
fibers and base spaces of different 
combinations with respect to their dimensions. 
In Appendix, we will give a partial description of general 
HJ theory for Nambu mechanics of $(n+1)$-dimensional 
phase space with $n$ Hamiltonians as 
a natural extension of the case $n=2$. 
In the present section, 
instead of discussing such formal extensions of our formalism, 
we present some concrete 
computations by taking the example 
of the Euler top, $G=\frac{1}{2}\Bigl(\frac{\xi_1^2}{I_1}+\frac{\xi_2^2}{I_2}+\frac{\xi_3^2}{I_3}\Bigr)$ and $H=\frac{1}{2}(\xi_1^2+\xi_2^2+\xi_3^2)$. It would help us to understand more deeply 
the meaning and working of our general constructions. 

Let us start from the (1/2)-formalism. By an N-gauge transformation 
as discussed in Sect. 3, 
we can replace $G$ by
\begin{align}
G-H/I_3\rightarrow G=\frac{\alpha}{2}(\xi^1)^2+
\frac{\beta}{2}(\xi^2)^2
\end{align}
where 
\begin{align}
\alpha=\frac{I_3-I_1}{I_3I_1}, \quad 
\beta=\frac{I_3-I_2}{I_3I_2}.
\end{align}
Thus the generalized HJ equations for the 1-form $\bar{\Omega}^{(1)}=S_{\mu}d\xi^{\mu}$ are reduced, with $S_i=-S\partial_iG$ and $S_0=0$, to
\begin{align}
\frac{\partial S}{\partial t}=-\bar{H}=-\frac{1}{2}\bigl(
(\xi^1)^2+(\xi^2)^2+(\xi^3)^2\bigr)
\end{align}
where
\begin{align}
\xi^3=\partial_1G\partial_2S-\partial_2G\partial_1S=
\alpha\xi^1\partial_2S-\beta\xi^2\partial_1S.
\end{align}
In view of the elliptical form of $G$, it is convenient to change the variables, $(\xi^1,\xi^2)
\rightarrow (G, u)$, 
\begin{align}
\xi^1=\sqrt{\frac{2G}{\alpha}}\sn u, \quad 
\xi^2=\sqrt{\frac{2G}{\beta}}\cn u,
\end{align}
by using 
Jacobi's elliptic functions,\footnote{Note that here we have used abbreviated notation 
for the elliptic functions $\sn (u;k), \cn (u; k)$ and $\dn (u;k)$ 
by suppressing implicit dependencies on the parameter $k$.}   satisfying
$
\sn^2 u+\cn^2 u=1, 
k^2\sn^2 u+\dn^2 u=1
$
and
$
\sn' u=\cn u\dn u, 
\cn' u=-\sn u\dn u, 
\dn' u=-k^2\sn u\cn u, 
$
where the modulus parameter $k$ is a constant to be fixed later 
such that the generalized HJ equation takes a simple form 
that is most convenient for integration. Of course, we should 
expect that when $\alpha=\beta$, corresponding to 
a symmetrical top, the above coordinate 
transformation would reduce to the usual 
polar coordinates with $k=0$. 
Then, 
\begin{align}
\xi^3=-\frac{\sqrt{\alpha\beta}}{\dn u}\partial_uS. 
\end{align}
By setting
\begin{align}
S=-E(t-t_0)+\bar{S}(\xi^1,\xi^2)
\end{align}
where $E$ and $t_0$ are constants, the equation is reduced to 
a time-independent one: 
\begin{align}\frac{\alpha\beta}{2\dn^2 u}(\partial_u\bar{S})^2=E-\frac{G}{\alpha}\sn^2 u-\frac{G}{\beta}\cn^2 u
=E+\frac{G(\alpha-\beta)}{\alpha\beta}\sn^2 u-\frac{G}{\beta}.
\end{align}
We fix the parameter $k$ by
\begin{align}
k^2=\frac{G(\beta-\alpha)}{\alpha\beta E_0}, 
\quad 
E_0=E-\frac{G}{\beta}, \nonumber 
\end{align}
which allows us to integrate $\bar{S}$ as
\begin{align}
\bar{S}=\sqrt{\frac{2E_0}{\alpha\beta}}\int_0^u
du \, \dn^2 u\equiv \sqrt{\frac{2E_0}{\alpha\beta}}=\frac{A}{\alpha\beta}
{\cal E}(u),
\end{align}
where we have chosen $\bar{S}|_{u=0}=0$ without 
losing generality and also redefined a constant $A=\sqrt{2\alpha\beta E_0}$ for later convenience. 
Thus $\xi^3$ is given by
\begin{align}
\xi^3=-\sqrt{2E_0}\dn u. 
\label{xi3}
\end{align}

The function ${\cal E}(u)$ is known as the 
fundamental elliptic integral of the second kind 
(or Jacobi's epsilon function; see, e.g., Ref. \cite{whittaker}. ) 
This function can also be expressed as
\begin{align}
{\cal E}(u)=\int_0^{\sn u}dx \sqrt{\frac{1-k^2x^2}{1-x^2}}, 
\label{E-integral}
\end{align}
which can easily be proven by noting that 
\[
\frac{d{\cal E}}{du}=\dn^2 u=\cn u \dn u\frac{\dn u}{\cn u}
=
\frac{\sqrt{1-k^2\sn^2 u}}{\sqrt{1-\sn^2 u}}\,\frac{d\sn u}{du}. 
\]
Thanks to this formula, we now have the desired form of  complete 
solution, which is expressed in terms of the original 
independent variables $(\xi^1,\xi^2,t)$:
\begin{align}
S=-E(t-t_0)+\frac{A}{\alpha\beta}
\int_0^{\xi^1/\sqrt{(\xi^1)^2+\frac{\beta}{\alpha}(\xi^2)^2}}
\sqrt{\frac{1-k^2x^2}{1-x^2}}dx. 
\end{align}

We can then follow the general prescription given in Sect. 3 
to derive the equations of motion. 
We first impose the Jacobi condition for the above solution 
with respect to the integration constant $E$:
\begin{align}
\frac{\partial S}{\partial E}={\rm const}. 
\end{align}
Since we have already introduced $t_0$ corresponding to the shift 
$S\rightarrow S+Et_0$, the constant on the r.h.side can actually be 
absorbed in $t_0$. According to our general formalism, 
this condition, together with the constraint
$
dG/dt=0, 
$
allows us to obtain the general solution 
of the Nambu equations of motion.  By taking a time-derivative, we obtain
\begin{align}
\frac{d}{dt}\Bigl(\frac{\partial \bar{S}}{\partial E}\Bigr)=\frac{du}{dt}\partial_u\Bigl(
\frac{\partial \bar{S}}{\partial E}\Bigr)=1.
\end{align}
It should be kept in mind that the derivative with respect to 
$E$ must be taken whil keeping $(\xi^1,\xi^2,t)$ fixed. Using 
$dA/dE=\alpha\beta/A$ and $dk^2/dE=-2\alpha\beta k^2/A^2$, 
\[
\frac{\partial \bar{S}}{\partial E}
=\frac{\alpha\beta}{A}
\frac{\partial \bar{S}}{\partial A}-
\frac{2\alpha\beta k^2}{A^2}\frac{\partial \bar{S}}{\partial k^2}
=\frac{1}{A}\Bigl({\cal E}-2k^2\frac{\partial {\cal E}}{
\partial k^2}\Bigr), 
\]
where, for the first equality, we treated 
$A$ and $k^2$ as independent variables. 
Using the integral representation \eqref{E-integral} and the properties of elliptic functions, we derive 
\begin{align}
\frac{d}{dt}\frac{\partial}{\partial k^2}{\cal E}=
-\frac{1}{2k^2}(1-\dn^2 u)\frac{du}{dt}
\label{Eformula}
\end{align}
and from the definition of ${\cal E}$ we also have
\[
\frac{d}{dt}{\cal E}=\dn^2 u\, \frac{du}{dt}.
\]
Putting these results together, we finally 
arrive at 
\[
\frac{du}{dt}=A, 
\]
and hence $u=At$ up to an arbitrary choice of the origin of time $t$. 
Substituting this result into the expressions for $(\xi^1,\xi^2,\xi^3)$ immediately gives a standard form of the general solution (see, e.g., Ref. \cite{MacM}) in terms of the elliptic functions, which is usually obtained 
by directly integrating the equations of motion 
using the conservation of $H$ and $G$. As a check 
of these results, we can derive, e.g., from Eq. \eqref{xi3}, 
\[
\frac{d\xi^3}{dt}=\sqrt{2E_0}Ak^2\sn u \cn u=\frac{I_1-I_2}{I_1I_2}\xi^1\xi^2. 
\]

Next let us treat the same system by applying the (1/1/1)-formalism. 
In this case, using $\partial_3H=\xi^3$, we first obtain 
an expression for $p_1$:
\begin{align}
p_1=-\int^{\xi^2}\frac{dx}{\sqrt{2E-(\xi^1)^2-x^2}}=-\arcsin \frac{\xi^2}{\sqrt{2E-(\xi^1)^2}}\nonumber 
\end{align}
or equivalently
\begin{align}
\xi^2=-\sqrt{2E-(\xi^1)^2} \sin(\partial_1T). 
\end{align}
We then have to solve
\begin{align}
\partial_tT=-\bar{\bar{G}}=-\frac{1}{2}(\alpha(\xi^1)^2+\beta(\xi^2)^2). 
\end{align}
By introducing a constant $F$ such that
\[
T=-Ft+\bar{T}(\xi^1)
\]
the equation is reduced to
\begin{align}
\sin \Bigl(\frac{d\bar{T}}{d\xi^1}\Bigr)=\sqrt{\frac{2F-\alpha(\xi^1)^2}{
\beta(2E-(\xi^1)^2)}}, 
\end{align}
and we obtain a complete solution
\begin{align}
\bar{T}=\int^{\xi^1}  \arcsin \sqrt{\frac{2F-\alpha x^2}{
\beta(2E-x^2)}}\, dx
\end{align}
with $F$ being the integration constant. 

The general solution for the trajectory $\xi^1(t)$ is 
derived by imposing the Jacobi condition
\begin{align}
-t_0&=\frac{\partial T}{\partial F}=-t +\frac{1}{\sqrt{2\alpha\beta(E-F/\beta)}}
\int^{\sqrt{\alpha}\xi^1/\sqrt{2F}}\frac{dx}{\sqrt{1-x^2}\sqrt{1-k^2x^2}}
\nonumber \\
&=-t+\frac{1}{\sqrt{2\alpha\beta(E-F/\beta)}}u\Bigl(\frac{\sqrt{\alpha}\xi^1}{\sqrt{2F}}\Bigr)
\end{align}
where the integration variable is 
rescaled $x\rightarrow \sqrt{2F}x/\sqrt{\alpha}$, and $u=u(x)$ is the inverse of the elliptic function: $x=\sn (u; k)$ 
with the modulus parameter $k^2=(\beta-\alpha) /\alpha(\beta E-F)$. 
We thus obtain
\begin{align}
\xi^1(t)=\sqrt{\frac{2F}{\alpha}}\sn (\sqrt{2\alpha\beta(
E-F/\beta)}(t-t_0); k), 
\end{align}
which coincides with the previous result of the (1/2)-formalism after 
renaming the integration constant as
$F\rightarrow G$. Other components are 
automatically satisfied by our general arguments.

Comparing with the well-known and traditional Hamilton-Jacobi treatments (see, e.g., Ref. \cite{MacM})  
of the Euler top in terms of Euler angles and the separation 
of variables, the new methods illustrated here on the basis 
of our generalized HJ theory of Nambu mechanics 
are much more direct and elegant, in the sense that 
the components of angular momentum themselves are 
treated as canonical coordinates.  

\section{Towards quantization}
\setcounter{equation}{0}

So far, we have not emphasized the relevance to our 
development of the canonical structure associated with 
the Nambu bracket. 
There is in fact a 
natural interpretation of our formulation of 
generalized HJ theory from the viewpoint of its connection with the 
Nambu bracket. It will lead us to a new standpoint towards quantization 
of Nambu mechanics.

As has already been pointed out 
in Ref.\cite{takh}, for any realization of the Nambu bracket 
satisfying the Fundamental identity (FI), we can define subordinated 
Poisson brackets that are intrinsically 
associated with the Nambu bracket. 
For example, 
if a function $G=G(\xi^1,\xi^2,\xi^3)$ is fixed, 
we can define
\begin{align}
\{A, B\}_G\equiv \{A, G, B\}. 
\label{subpoisson}
\end{align}
The Jacobi identity is automatically satisfied 
because of the FI,
\begin{align}
&\{A,G,\{B,F,C\}\}\nonumber \\
&=\{\{A,G,B\},F,C\}
+\{B,\{A,G,F\},C\}+\{B,F,\{A,G,C\}\}
\end{align}
which reduces to the Jacobi identity for \eqref{subpoisson} 
by setting $F=G$. If we assume 
$\partial_3G=0$, we have 
\begin{align}
\{\xi^1,\xi^2\}_G=0, \quad 
\{\xi^3, \xi^1\}_G=-\partial_2G, \quad 
\{\xi^3, \xi^2\}_G=\partial_1G, 
\label{Gpoisson}
\end{align}
and the Nambu equations of motion take the standard 
Hamiltonian form, 
\begin{align}
\frac{d\xi^i}{dt}=\{H, \xi^i\}_G. 
\end{align}
This shows that the (1/2)-formalism with $\partial_3G=0$ 
can be regarded as the classical limit of a 
quantized Nambu mechanics; it is 
obtained by introducing a wave function that we denote by 
$\langle \xi^1,\xi^2|1(t)\rangle $ in the representation where $\xi^1, \xi^2$ are 
diagonalized, corresponding to the first component of Eq. \eqref{Gpoisson}, and $\xi^3$ is replaced by 
a differential operator acting on the wave function, 
\begin{align}
\xi^3\rightarrow -i\hbar (\partial_1G\partial_2-\partial_2G\partial_1)
=-i\hbar d_G, 
\end{align}
which is consistent with the above subordinated Poisson brackets 
(the last two of them) 
and the generalized HJ equations in the reduced 
form given in Sect. 3.3, if we assume 
the usual relationship between Poisson brackets and 
commutators. The Schr\"{o}dinger equation is then 
\begin{align}
i\hbar\partial_t\langle \xi^1,\xi^2|1(t)\rangle=H(\xi^1,\xi^2, -i\hbar d_G)\langle \xi^1,\xi^2|1(t)\rangle,
\end{align}
which leads us to the (1/2)-formalism 
under an ansatz $\langle \xi^1,\xi^2|1(t)\rangle\sim e^{iS(\xi^1,\xi^2,t)/\hbar}$. 
The ket symbol $|1(t)\rangle$ is meant to imply that 
 it contains information on 
the initial conditions corresponding classically to 
a single integration constant $Q_1$. 

Similarly, if we choose another Poisson bracket by exchanging 
$G$ for $H$ and assuming $\partial_3H\ne 0$, 
\begin{align}
\{A,B\}_H\equiv \{A, H,B\},
\end{align}
we have 
\begin{align}
\{\xi^1,\xi^2\}_H=-\partial_3H, \quad \{\xi^3, \xi^1\}_H=-\partial_2H, 
\quad \{\xi^3, \xi^2\}_H=\partial_1H. 
\label{bra1/1/1}
\end{align}
Then, the first bracket 
can be interpreted as defining the symplectic 2-form 
$d\xi^1\wedge d\xi^2/\partial_3H$ which we have defined in the 
phase space $(\xi^1,\xi^2, t)$ of the (1/1/1)-formalism, 
under the constraint $H=E$: the latter constraint, 
in principle, enables us to express $\xi^3=\xi^3(\xi^1, \xi^2; E)$ in terms of 
$\xi^1$ and $\xi^2$. Indeed, once the first bracket \eqref{bra1/1/1} is 
given, the remaining two are consequences of this constraint. 
This is due to the 
following identities:
\begin{align}
&0=\partial_2H(\xi^1,\xi^2,\xi^3(\xi^1,\xi^2;E))=\partial_2H+
\partial_3H\partial_2\xi^3, \nonumber \\
&0=\partial_1H (\xi^1,\xi^2,\xi^3(\xi^1,\xi^2;E))=
\partial_1H+\partial_3H\partial_1\xi^3. 
\end{align}
In this case, 
the wave function is a function of $(\xi^1,t)$, denoted by 
$\langle \xi^1|2(t)\rangle $ and the Schr\"{o}dinger equation is 
\begin{align}
i\hbar\partial_t\langle \xi^1|2(t)\rangle=\bar{G}(\xi^1, \hat{\xi}^2)\langle \xi^1|2(t)\rangle,
\end{align}
where $\hat{\xi}^2$ is a differential operator, formally given by 
\begin{align}
-i\hbar\partial_1=-\int^{\hat{\xi}^2}
\frac{dx}{\partial_3H(\xi^1, x)}. 
\end{align}
The (1/1/1)-formalism is then obtained in 
the classical limit with an ansatz $\langle \xi^1|2(t)\rangle\sim e^{iT(\xi^1,t)/\hbar}$. The ket $|2(t)\rangle$ is meant to imply 
two integration constants $(Q_1,Q_2)$ classically as initial conditions. 

Rigorously speaking, these formal 
constructions (especially, the second one) of quantum theory are not in general well defined as they stand, due to the ambiguity 
of operator orderings, 
since both $H$ and $G$ can be arbitrarily 
complicated functions. However, they are 
meaningful at least in a semiclassical limit. 
More importantly, these quantized 
theories, which seem to be 
entirely different from each other with different Hilbert spaces 
of wave functions and 
different Schr\"{o}dinger (or Heisenberg) equations,  are 
guaranteed to give one and the same Nambu equations of motion 
in the classical limit. Classically, they are 
connected by N-gauge transformations and/or 
canonical coordinate transformations in the sense 
of the original phase space $(\xi^1,\xi^2,\xi^3,t)$. 
From this viewpoint, one possible standpoint 
towards more general and rigorous formulations of the 
quantum theory of Nambu mechanics seems to
\begin{enumerate}
\item[(1)] quantize the subordinated Poisson 
brackets defined by usual commutators algebras with 
variable choices of two Hamiltonians $(H,G)$ with respect 
to N-gauge transformations;
\item[(2)] enlarge the usual framework of quantum mechanics to 
a new extended scheme, allowing (infinitely) many different 
Hilbert spaces corresponding to different 
choices of Poisson brackets and different 
Hamiltonians;
\item[(3)] construct a transformation theory by 
which we can transform systems among the sets of Hilbert spaces and corresponding 
Hamiltonians in some covariant fashion,  
such that it gives the N-gauge and 
canonical coordinate transformations 
in the classical limit;
\item[(4)] find probabilistic 
interpretations of the formalism, in such a manner that 
different and allowed choices of Hilbert spaces and 
Hamiltonians in the framework of transformation theory give physically unique results. 
\end{enumerate}
The most challenging and imaginative parts of this program 
would be (3) and (4), by which we should expect that
 various possible ambiguities associated with transition from classical theory to 
quantum theory would be removed or restricted 
appropriately. 

We emphasize that the above program is certainly a possible 
route towards quantization, though, to the author's 
knowledge, such a viewpoint 
has scarcely
 been stressed in the literature.\footnote{For a review on quantization and related matters, we refer the interested 
 readers to \cite{n-ary} which 
contains an extensive list of 
references. It is also to be noted that an example of quantized Nambu mechanics based on the reduced Poisson bracket \eqref{bra1/1/1} is 
discussed in \cite{axenides}. The present author 
thanks to the authors of the last reference for 
bringing their works to his attention. } We hope that this new viewpoint and our generalized 
HJ theory would be useful ultimately 
for further applications of Nambu mechanics and the Nambu bracket 
in the arena of fundamental physics.

\vspace{0.5cm}
\noindent
{\large Acknowledgements}

The author would like to dedicate this paper to 
the memory of Yoichiro Nambu. Many of his works 
and also occasional conversations with him have been 
sources of inspiration to the present author. 

The present work is supported in part by Grant-in-Aid for 
Scientific Research (No. 25287049) from the Ministry of 
Education, Science, and Culture. 


\vspace{0.5cm}
\noindent
{\large Appendix: Generalilzed HJ formalism for Nambu mechanics of order $n$}
\renewcommand{\theequation}{A.\arabic{equation}}
\appendix 
\setcounter{equation}{0}
\vspace{0.2cm}
\noindent

Here we extend the 
$(1/2)$-formalism of Sect. 3 to a $(1/n)$-formalism 
in the case of $(n+1)$-dimensional phase space $(\xi^1, \ldots, 
\xi^{n+1})$. 
We denote one of $n$ Hamiltonians by $H$ and 
the remaining ones by $G_a$ $(a=1, \ldots, n-1)$. 
The equations of motion are
\begin{align}
\frac{d\xi^i}{dt}=\{H, G_1, \ldots, G_{n-1}, \xi^i\}
\equiv X^i.
\end{align}
The EE equations are obtained from 
an $n$-form
\begin{align}
\Omega^{(n)}=(-1)^n\xi^{n+1}d\xi^1\wedge \cdots \wedge 
d\xi^n-HdG_1\wedge \cdots \wedge dG_{n-1}\wedge dt
\end{align}
as the vanishing condition $d\bar{\Omega}^{(n)}=0$ 
after the projection $\Omega^{n}\rightarrow 
\bar{\Omega}^{n}$ by setting $\xi^{n+1}=\xi^{n+1}
(\xi^1,\ldots, \xi^n,t)$, 
corresponding to the property (Ref. \cite{takh}) that 
the vector differential operator $\tilde{X}=X^i\partial_i+\partial_t$ is 
a line field for the exact $(n+1)$-form $\Omega^{(n+1)}\equiv d\Omega^n$. 
The generalized HJ equations are then obtained 
by demanding
\begin{align}
\bar{\Omega}^{n}=dS^{(n-1)}
\end{align}
for an $(n-1)$-form $S^{(n-1)}$ on the base space, 
\begin{align}
S^{(n-1)}=&\frac{1}{(n-1)!}\sum_{i_1,\ldots,i_{n-1}=1}^{n}
S_{i_1,\ldots, i_{n-1}}d\xi^{i_1}\wedge \cdots \wedge d\xi^{i_{n-1}}
\nonumber\\
&+\frac{1}{(n-2)!}\sum_{i_1,\ldots,i_{n-2}=1}^n
S_{i_1,\ldots, i_{n-2},0}d\xi^{i_1}\wedge \cdots \wedge d\xi^{i_{n-2}}
\wedge dt, 
\end{align}
where $S_{i_1,\ldots, i_{n-1}}$ and $S_{i_1,\ldots, i_{n-2},0}$ 
are 
completely antisymmetric tensors with respect to 
spatial indices: 
\begin{align}
-\bar{H}&\frac{\partial(\bar{G}_1,\ldots, \bar{G}_{n-1})}{\partial
(\xi^{i_1}, \ldots, \xi^{i_{n-1}})}=(-1)^{n-1}\bigl(\partial_t S_{i_1,\ldots, i_{n-1}}-
\frac{1}{(n-2)!}\sum_{P(i_1,\ldots, 
i_{n-1})}(-1)^{\epsilon(P)}\partial_{i_{n-1}}S_{i_1,\ldots, i_{n-2},0}\bigr),  \\
&\xi^{n+1}=-\frac{1}{(n-1)!}\sum_{i_1,\ldots,i_n=1}^{n}\Bigl(\sum_{P(i_1,\ldots,  i_n)}(-1)^{\epsilon(P)}\partial_{i_n}S_{i_1, \ldots, i_{n-1}}\Bigr), 
\end{align}
where, as in the case of $n=2$, we defined projected Hamiltonians 
such as  
\begin{align}
\bar{H}=H\bigl(\xi^1,\ldots, \xi^n, \xi^{n+1}(\xi^1,\ldots, \xi^n,t)\bigr). 
\end{align}
Also note that $(-1)^{\epsilon(P)}$ is the parity of permutations $P$ of the set of 
indices $(1,2,\ldots, n)$ as 
indicated for summation symbols: depending on 
even or odd permutations, $\epsilon(P)=0$ or $1$.  
The S-gauge transformations as a symmetry 
of this set of equations are, with $\lambda_{i_1,\ldots, i_{n-2}}$ 
being an arbitrary completely antisymmetric $(n-2)$-tensor, 
\begin{align}
&S_{i_1,\ldots, i_{n-1}}\rightarrow S_{i_1,\ldots, i_{n-1}}
-\frac{1}{(n-2)!}\sum_{P(i_1,\ldots, 
i_{n-1})}(-1)^{\epsilon(P)}
\partial_{i_{n-1}}\lambda_{i_1,\ldots, i_{n-2}}, \\
&S_{i_1,\ldots, i_{n-2},0}\rightarrow S_{i_1,\ldots, i_{n-2},0}
-\partial_t\lambda_{i_1,\ldots, i_{n-2}}.
\end{align}
In what follows, we choose the gauge condition 
\begin{align}
S_{i_1,\ldots, i_{n-2},0}=0. 
\end{align}
On the other hand, 
the N-gauge symmetry of the equations of motion 
is generated by transformations $(H,G_1, \ldots, G_{n-1})
\rightarrow (H', G'_1, \ldots, G'_{n-1})$ of Hamiltonians such that 
\begin{align}
\frac{\partial (H', G'_1, \ldots, G'_{n-1})}{\partial(H,G_1, \ldots, G_{n-1})}=1. 
\end{align}

As in the (1/2)-formalism, we first assume, utilizing the 
N-gauge symmetry and/or canonical coordinate 
transformations, that 
we can choose such that the $G_a$ do not depend on 
$\xi^{n+1}$: 
\begin{align}
\partial_{n+1}G_a=0. 
\end{align}
Then, by using $\partial_t\bar{G}_a=\partial_tG_a=0$ and defining a scalar $S$, 
\begin{align}
S_{i_1,\ldots, i_{n-1}}=(-1)^{n-1}\frac{\partial(G_1,
\ldots, G_{n-1})}{\partial(\xi^{i_1},\ldots, \xi^{i_{n-1}})}S
\end{align}
the generalized HJ equations are 
reduced to
\begin{align}
&\partial_tS=-H(\xi^1,\ldots, \xi^n, \xi^{n+1}),\nonumber 
\\
&\xi^{n+1}=
\frac{(-1)^{n}}{(n-1)!}\sum_{P(i_1,\ldots, i_n)}(-1)^{\epsilon(P)}
\frac{\partial(G_1,\ldots, G_{n-1})}{\partial(\xi^{i_1},
\ldots,  \xi^{i_{n-1}})}\partial_{i_n}S. 
\end{align}

To make things more concrete, let us collect some relevant expressions for 
$n=3$, since the above 
general expressions are not very elegant to deal with. 
The 2-form is 
\begin{align}
S^{(2)}=&S_{12}d\xi^1\wedge d\xi^2+S_{23}d\xi^2\wedge d\xi^3
+S_{31}d\xi^3\wedge d\xi^1\nonumber \\
&+S_{10}d\xi^1\wedge dt+S_{20}d\xi^2\wedge dt +S_{30}d\xi^3
\wedge dt.
\end{align}
The generalized HJ equations are
\begin{align}
-\bar{H}&\frac{\partial(\bar{G}_1,\bar{G}_2)}{\partial(\xi^1,\xi^2)}=\partial_tS_{12}
-\partial_2S_{10}+\partial_1S_{20},\label{A16}\\
-\bar{H}&\frac{\partial(\bar{G}_1,\bar{G}_2)}{\partial(\xi^2,\xi^3)}=\partial_tS_{23}
-\partial_3S_{20}+\partial_2S_{30},\label{A17}\\
-\bar{H}&\frac{\partial(\bar{G}_1,\bar{G}_2)}{\partial(\xi^3,\xi^1)}=
\partial_tS_{31}-\partial_3S_{10}+\partial_1S_{30},
\label{A18}\\
&-\xi^4=\partial_3S_{12}+\partial_1S_{23}+\partial_2S_{31}. 
\label{A19}
\end{align}
The S-gauge transformations are 
\begin{align}
S_{ij}\rightarrow S_{ij}+\partial_i\lambda_j-\partial_j\lambda_i, 
\quad S_{i0}\rightarrow S_{i0}-\partial_t\lambda_i.
\end{align}
The generalized HJ equations under the assumption $\partial_4G_1=\partial_4G_2=0$ and the gauge condition $S_{i0}=0$ reduce to
\begin{align}
\partial_tS+H(\xi^1,\xi^2,\xi^3, \xi^4)=0, 
\end{align}
\begin{align}
\xi^4=-\partial_1S_{23}-\partial_2S_{31}-\partial_3S_{12}
=-\frac{\partial(G_1,G_2,S)}{\partial(\xi_1,\xi_2,\xi_3)}\equiv -
d_GS
\label{A22}
\end{align}
with
\begin{align}
S_{ij}=\frac{\partial(G_1,G_2)}{\partial(\xi_i,\xi_j)}S, 
\end{align}
where we used 
\[
\partial_1\frac{\partial(G_1,G_2)}{\partial(\xi^2,\xi^3)}
+\partial_2\frac{\partial(G_1,G_2)}{\partial (\xi^3,\xi^1)}
+\partial_3\frac{\partial(G_1,G_2)}{\partial (\xi^1,\xi^2)}=0. 
\]

For this case, let us consider the converse problem, namely, 
the derivation of the equations of motion from these equations. 
We first consider the derivation of the 
equation of motion for $\xi^4$ from the integrability 
condition for our generalized HJ equations. Taking a partial 
derivative of \eqref{A19} with respect to time and 
using \eqref{A16}$\sim$\eqref{A18}, we obtain
\begin{align}
&\partial_t\xi^4=\partial_3[\bar{H}(\partial_1G_1\partial_2G_2-\partial_2G_1\partial_1G_2)]\nonumber \\
&+\partial_1[\bar{H}(\partial_2G_1\partial_3G_2-\partial_3G_1\partial_2G_2)]
+\partial_2[\bar{H}(\partial_3G_1\partial_1G_2-\partial_1G_1\partial_3G_2)]\nonumber 
\\
&=\partial_1\bar{H} (\partial_2G_1\partial_3G_2-\partial_3G_1\partial_2G_2)
+\partial_2\bar{H} (\partial_3G_1\partial_1G_2-\partial_1G_1\partial_3G_2)
\nonumber \\
&+\partial_3\bar{H}(\partial_1G_1\partial_2G_2-\partial_2G_1\partial_1G_2),
\end{align}
where the partial derivatives with respect to $(\xi^1,\xi^2,\xi^3)$ 
are performed by regarding $\xi^4$ involved in $H$ as 
a function of them. To convert this result into 
the equations in the phase space, 
we have to rewrite their partial derivatives 
by treating $\xi^4$ as an independent variable. 
Using \eqref{A22}, we have 
\begin{align}
&\partial_i\bar{H}=\partial_iH+\partial_4H\partial_i\xi^4
=\partial_iH-\partial_4H\partial_i(\partial_1S_{23}+\partial_2S_{31}
+\partial_3S_{12}),\\
&\partial_t\xi^4=\frac{d\xi^4}{dt}-\sum_{i=1}^3\partial_i\xi^4
\frac{d\xi_i}{dt}=\frac{d\xi^4}{dt}+
\sum_{i=1}^3\frac{d\xi^i}{dt}\partial_i(\partial_1S_{23}+\partial_2S_{31}
+\partial_3S_{12}).
\end{align}
Let us first assume the equations of motion for 
$\xi_i$ $(i=1,2,3)$:
\begin{align}
&\frac{d\xi_1}{dt}=\{H, G_1,G_2,\xi_1\}=-
\partial_4H(\partial_2G_1\partial_3G_2-\partial_3G_1\partial_2G_2)
=-\partial_4H\frac{\partial(G_1,G_2)}{\partial(\xi_2,\xi_3)},\\
&\frac{d\xi_2}{dt}=\{H, G_1,G_2,\xi_2\}=
\partial_4H(\partial_1G_1\partial_3G_2-\partial_3G_1\partial_1G_2)
=-\partial_4H\frac{\partial(G_1,G_2)}{\partial(\xi_3,\xi_1)},\\
&\frac{d\xi_3}{dt}=\{H, G_1,G_2,\xi_3\}=-
\partial_4H(\partial_1G_1\partial_2G_2-\partial_2G_1\partial_1G_2)
=-\partial_4H\frac{\partial(G_1,G_2)}{\partial(\xi_1,\xi_2)}.
\end{align}
Substituting these expressions into the 
integrability condition, we find that the 
terms proportional to $\partial_4H$ on both sides 
just cancel each other, and 
the final result is, as promised,  
\begin{align}
\frac{d\xi_4}{dt}=
\partial_1H\frac{\partial(G_1,G_2)}{\partial(\xi_2,\xi_3)}
+\partial_2H\frac{\partial(G_1,G_2)}{\partial(\xi_3,\xi_1)}
+\partial_3H\frac{\partial(G_1,G_2)}{\partial(\xi_1,\xi_2)}
=\{H,G_1,G_2,\xi_4\}. 
\end{align}

Next we have to 
derive the equations of motion for the base coordinates $(\xi_1,\xi_2,\xi_3)$ themselves, 
by imposing Jacobi-type conditions appropriately. 
As a generalization of the case $n=2$, 
we suppose that we are given a complete solution that has  three integration constants, and 
pick up one of them, which is denoted by $Q$. 
Then we demand that the derivative $\frac{\partial S}{\partial Q}$ is a 
constant that is independent of time, 
\begin{align}
0=\frac{d}{dt}\frac{\partial S}{\partial Q}=\frac{\partial^2 S}{\partial Q\partial t}+\sum_{i=1}^3\frac{\partial^2 S}{\partial\xi^i\partial Q}\frac{d\xi^i}{dt}, 
\label{jacobi-appen}
\end{align}
together with the conditions
\begin{align}
0=\frac{dG_1}{dt}=\sum_{i=1}^3\frac{\partial G_1}{\partial\xi^i}\frac{d\xi^i}{dt}, \quad 
0=\frac{dG_2}{dt}=\sum_{i=1}^3\frac{\partial G_2}{\partial\xi^i}\frac{d\xi^i}{dt}.
\end{align}
We have to require that 
the choice of $Q$ is such that these three conditions enable 
us to solve $(\xi_1,\xi_2,\xi_3)$ as functions of time $t$. 
Now using the generalized HJ equations, we have
\begin{align}
\frac{\partial^2 S}{\partial Q\partial t}=-\frac{\partial H}{\partial \xi^4}
\frac{\partial \xi^4}{\partial Q}=\partial_4H\Bigl(
\frac{\partial(G_1,G_2)}{\partial(\xi^1,\xi^2)}\frac{
\partial^2S}{\partial \xi^3\partial Q}+\frac{\partial(G_1,G_2)}{\partial(\xi^2,\xi^3)}\frac{
\partial^2S}{\partial \xi_1\partial Q}+
\frac{\partial(G_1,G_2)}{\partial(\xi^3,\xi^1)}\frac{
\partial^2S}{\partial \xi^2\partial Q}\Bigr).
\end{align}
Thus the condition \eqref{jacobi-appen} is rewritten as
\begin{align}
0=\sum_{i=1}^3
\frac{\partial^2 S}{\partial \xi^i\partial Q}
\Bigl(\frac{d\xi^i}{dt}+\partial_4H
\frac{\partial(G_1,G_2)}{\partial(\xi^j, \xi^k)}
\Bigr),
\label{A34}
\end{align}
where the set of indices $(i, j, k)$ should be 
understood as a cyclic permutation of the 
ordered set of indices $(1,2,3)$.

In order to be able to conclude from \eqref{A34} that 
it gives the equations of motion,  we further require that 
the two conditions for the conservation of $G_1,G_2$ are independent 
of each other. This implies that, out of three 
time derivatives $d\xi^1/dt, d\xi^2/dt, d\xi^3/dt$, we can solve two of them in terms of a single one. 
\begin{align}
\begin{pmatrix}
\partial_1 G_1 & \partial_2 G_1 \\
\partial_1 G_2 & \partial_2 G_2
\end{pmatrix}
\begin{pmatrix}
\dot{\xi}^1\\
\dot{\xi}^2
\end{pmatrix}=-\begin{pmatrix}
\partial_3 G_1\\
\partial_3 G_2
\end{pmatrix}\dot{\xi}^3.
\end{align}
Thus
\begin{align}
\begin{pmatrix}
\dot{\xi}^1\\
\dot{\xi}^2
\end{pmatrix}
=\Delta^{-1}\begin{pmatrix}
\partial_2G_2 &-\partial_2G_1 \\
-\partial_1G_2 & \partial_1G_1
\end{pmatrix}
\begin{pmatrix}
\partial_3G_1\\
\partial_3G_2
\end{pmatrix}
\dot{\xi}^3, 
\label{A37}
\end{align}
where
\begin{align}
\Delta=\partial_1G_1\partial_2G_2-
\partial_2G_1\partial_1G_2=\frac{\partial(G_1,G_2)}{\partial(\xi^1,\xi^2)}.
\end{align}
Then we substitute this result into the above 
condition \eqref{A34} and obtain
\begin{align}
0=&\frac{\partial^2 S}{\partial \xi^3\partial Q}
\Bigl(\frac{d\xi^3}{dt}+\partial_4H
\frac{\partial(G_1,G_2)}{\partial(\xi^1, \xi^2)}
\Bigr)
\nonumber \\
&+\frac{\partial^2 S}{\partial \xi^1\partial Q}
\Bigl(\Delta^{-1}(\partial_2G_2\partial_3G_1-\partial_2G_1\partial_3G_2)
\frac{d\xi^3}{dt}
+\partial_4H
\frac{\partial(G_1,G_2)}{\partial(\xi^2, \xi^3)}
\Bigr)\nonumber \\
&
+\frac{\partial^2S}{\partial\xi^2\partial Q}
\Bigl(\Delta^{-1}(-\partial_1G_2\partial_3G_1+\partial_1G_1\partial_3G_2)
\frac{d\xi^3}{dt}
+\partial_4H
\frac{\partial(G_1,G_2)}{\partial(\xi^3, \xi^1)}
\Bigr)
\nonumber \\
&=\Delta^{-1}
\Bigl(
\sum_{i=1}^3\frac{\partial^2 S}{\partial Q\partial \xi^i}
\frac{\partial(G_1,G_2)}{\partial(\xi^j, \xi^k)}\Bigr)
\Bigl[
\frac{d\xi_3}{dt}+\partial_4H\frac{\partial(G_1, G_2)}{\partial(\xi^1,\xi^2)}
\Bigr].
\end{align}
The expression appearing in the round bracket as the 
coefficient in the last line does not vanish if we require that $\partial S/\partial Q$ is not $d_G$ invariant: 
\begin{align}
d_G\Bigl(\frac{\partial S}{\partial Q}\Bigr)=\sum_{i=1}^3\frac{\partial^2 S}{\partial Q\partial \xi^i}
\frac{\partial(G_1,G_2)}{\partial(\xi^j, \xi^k)}\ne 0.
\end{align}
This requirement is sufficient to ensure the equation 
of motion for $\xi^3$, 
\begin{align}
\frac{d\xi^3}{dt}=-\partial_4H\frac{\partial(G_1, G_2)}{\partial(\xi^1,\xi^2)}, 
\end{align}
which also implies immediately the equations of motion 
for $\xi^1, \xi^2$ using the above formula \eqref{A37}, 
expressing $\dot{\xi}^1, \dot{\xi}^2$ in terms of $\dot{\xi}^3$. 
Thus we conclude that sufficient 
conditions for deriving equations of motion are 
\begin{align}
\Delta\ne0, 
\end{align}
and the requirement that $\partial S/\partial Q$ is not 
$d_G$ invariant. It is to be noted that if $\Delta=0$ 
we can turn to other choices of independent $\dot{\xi_i}$. 
Hence in general it is sufficient to 
require at least one of 
\[
\frac{\partial(G_1,G_2)}{\partial(\xi^i, \xi^j)}
\]
is nonzero out of the three possible ones. 

These conditions are naturally extended to general $n$. 
One is 
\begin{align}
\sum_{P(i_1, \ldots,  i_n)}(-1)^{\epsilon(P)}\frac{\partial^2 S}{\partial Q
\partial\xi_{i_n}}\frac{\partial(G_1,\ldots,G_{n-1})}{
\partial(\xi_{i_1},\ldots, \xi_{i_{n-1}})}\ne 0
\end{align}
and also that 
\begin{align}
\frac{\partial(G_1,G_2,\ldots,G_{n-1})}
{\partial(\xi_{i_1},\xi_{i_2}, \ldots, \xi_{i_{n-1}})}\ne 0
\end{align}
for at least one of the possible combinations of the set  
$(i_1,\ldots, i_{n})$ of indices. 

Finally, let us briefly consider an extension 
of the (1/1/1)-formalism to the general case: $(1/n)\rightarrow 
(1/1/n-1)$. We first set the condition 
\begin{align}
\bar{H}(\xi^1,\ldots, \xi^n)\equiv H(\xi^1,\ldots, \xi^n, \xi^{n+1}(\xi^1, \ldots, \xi^n))=E
\end{align}
 with a constant $E$, which implies
\begin{align}
1=\partial_{n+1}H\frac{\partial \xi^{n+1}}{\partial E}, 
\quad 
0=\partial_iH+\partial_{n+1}H\partial_i\xi^{n+1}, \quad 
(i=1, \ldots, n).
\end{align}
As has always been the case, for the partial derivatives 
without bar symbols it should be understood 
that $\xi^{n+1}=\xi^{n+1}(\xi^1,\ldots, \xi^n, t)$ is 
substituted after differentiation. 
Then, we have 
a completely time-independent solution to the generalized 
HJ equations: 
\begin{align}
&S_{i_1,\ldots, i_{n-1}}=g_{i_1,\ldots, i_{n-1}},\\
&S_{i_1,\ldots, i_{n-2},0}=(-1)^{n-1}EG_1
\frac{\partial(G_2,\ldots, G_{n-1})}{\partial(\xi^{i_1},\ldots,
\xi^{i_{n-2}})}, \\
&\xi^{n+1}=-\frac{1}{(n-1)!}
\sum_{i_1,\ldots, i_n=1}^n
\Bigl(\sum_{P(i_n,i_1,\ldots, i_{n-1})}(-1)^{\epsilon(P)}
\partial_{i_n}g_{i_1,\ldots, i_{n-1}}
\Bigr), 
\end{align} 
where $g_{i_1,\ldots, i_{n-1}}$ is a time-independent 
and completely antisymmetric $(n-1)$-tensor. 
The disappearance of time motivates us to decompose the 
$n$-dimensional base space into a fiber bundle with $(n-1)$-dimensional 
base space and one-dimensional fiber parametrized by $\xi^n$, and 
define 
a closed (and exact) $n$-form $\bar{\Omega}'^{(n)}$ by regarding the base space
 $(\xi^1,\ldots, \xi^n, t)$ now as an $n$-dimensional phase space: 
\begin{align}
\bar{\Omega}'^{(n)}=(-1)^n\frac{\partial\xi^{n+1}}{\partial E}
d\xi^1\wedge \cdots \wedge d\xi^n-
d\bar{G}_1\wedge \cdots d\bar{G}_{n-1}\wedge dt=d\Omega'^{n-1}, 
\label{closedbarnform}
\end{align}
with
\begin{align}
\Omega'^{(n-1)}=-\Bigl(\int^{\xi^n}
\frac{dx}{\partial_{n+1}H(\xi^1,\ldots,\xi^{n-1}, x)}
\Bigr)d\xi^1\wedge \cdots \wedge d\xi^{n-1}
-\bar{G}_1d\bar{G}_2\wedge \cdots \wedge d\bar{G}_{n-1}\wedge dt. 
\end{align}

On the other hand, 
the Nambu equations of motion for $\xi^i$ $(i=1,\ldots, n)$ are rewritten as 
\begin{align}
\frac{d\xi^i}{dt}=(-1)^n\partial_{n+1}H
\{\bar{G}_1,\ldots, \bar{G}_{n-1}, \xi^i\}_{n}\equiv \bar{X}^i, 
\end{align}
where $\{\, \cdots \, \}_n$ denotes the $n$-dimensional 
Nambu bracket with respect to $(\xi^1,\ldots, \xi^n)$, and use has been made of 
\[
\partial_iG_a=\partial_i\bar{G}_a-\partial_i\xi^{n+1}\partial_{n+1}G_a=
\partial_i\bar{G}_a+
\frac{\partial_iH}{\partial_{n+1}H}\partial_{n+1}G_a. 
\]
The vector differential operator $\tilde{\bar{X}}=\bar{X}^i\partial_i+\partial_t$ 
is a null vector for the closed $n$-form \eqref{closedbarnform}. 
Correspondingly, 
the EE equations, as partial differential equations 
in $(n-1)$-dimensional (+time) 
base space $(\xi^1,\ldots, \xi^{n-1},t)$, are obtained as the vanishing condition 
$0=\bar{\bar{\Omega}}'^{n}\equiv \bar{\Omega}'^{n}|_{\xi^{n}=
\xi^{n}(\xi^1,\ldots, \xi^{n-1},t)}$. Similarly, reduced generalized 
HJ equations are obtained 
by 
requiring that the $(n-1)$ form $\Omega'^{(n-1)}$ is equal to 
the exterior derivative 
$d\Omega'^{(n-2)}$ of an $(n-2)$-form $\Omega'^{(n-1)}$ 
after  
the same projection to $(\xi^1,\ldots, \xi^{n-1},t)$. 
In this way, we can, in principle, continue 
reductions to lower-dimensional base spaces, recursively: 
 we have a ``nested" structure of generalized 
 HJ formalisms,  ranging all the way from 
$(1/n), (1/1/n-1),$ up to $(1/1/\ldots/1)$.


\end{document}